\documentclass[aps,twocolumn,pra]{revtex4-2}
\usepackage[english]{babel}

\usepackage{amsthm}
\usepackage{amsmath,bm,dsfont}
\usepackage{amssymb}
\usepackage{amsfonts}
\usepackage{graphicx}
\usepackage{physics}
\usepackage{txfonts}

\usepackage[colorlinks=true,linkcolor=blue,citecolor=blue,urlcolor=blue]{hyperref}

\begin{document}

\title{Squeezing-type lower bounds on generalized metrological sensitivity}

\author{Irene García-Martínez}
\email{irene.garcia@uv.es}
\affiliation{Instituto de Física Corpuscular (IFIC), CSIC‐Universitat de València and Departament de Física Teòrica, UV, Avinguda Vicent Andrés Estellés 19, E-46100 Burjassot (Valencia), Spain} 

\author{Manuel Gessner}
\email{manuel.gessner@uv.es}
\affiliation{Instituto de Física Corpuscular (IFIC), CSIC‐Universitat de València and Departament de Física Teòrica, UV, Avinguda Vicent Andrés Estellés 19, E-46100 Burjassot (Valencia), Spain} 

\begin{abstract}
We introduce squeezing-type lower bounds for a hierarchy of variance bounds under specific unbiasedness constraints, which generalizes the quantum Cramér-Rao bound. Our construction directly extends standard squeezing coefficients to the corresponding generalized metrological sensitivities while retaining their experimentally accessible structure. We show how observables maximizing these bounds can be identified analytically within families of accessible observables. If arbitrary observables are accessible, the optimized squeezing-type bound recovers the exact generalized metrological sensitivity. These bounds provide efficient and experimentally viable characterizations of generalized metrological sensitivities that are especially relevant in finite-shot scenarios and in the transition from local to global parameter estimation.
\end{abstract}

\maketitle
\section{Introduction}

Quantum limits on parameter estimation in the form of lower bounds on the variance of unbiased estimators define fundamental constraints on measurement precision of any physical parameter~\cite{helstrom1976,giovannetti2006,paris2009quantum,toth2014,pezze2018,braun2018,gessner2023hierarchies}. Their inverses yield metrological sensitivities that encode information-geometric structure~\cite{Braunstein1994,gessner2018,sidhu2020geometric} and may reveal nonclassical properties of quantum states, such as their multipartite entanglement structure~\cite{pezze2009,toth2012multipartite,hyllus2012fisher,ren2021young,szalay2025,smerzi2026architectures}. While families of fundamental bounds are known in principle, their exact evaluation and experimental saturation are often impractical, which motivates the use of efficiently computable and experimentally accessible lower bounds.

The most widely known fundamental limit in this context is the quantum Cramér–Rao bound (QCRB), whose inverse defines the quantum Fisher information (QFI), the most prominent and widely used measure of metrological sensitivity~\cite{helstrom1976,Braunstein1994,paris2009quantum,pezze2018}. Beyond its central role in quantum metrology, the QFI has found numerous applications in quantum information science, ranging from entanglement detection to information geometry~\cite{toth2014}. It can be approximated by squeezing coefficients~\cite{wineland1992spin,kitagawa1993,wineland1994squeezed}, which provide lower bounds obtained by restricting to the measurement of expectation values of linear observables, such as collective spin components or field quadratures. Since these bounds depend only on first and second moments, they can be interpreted as Gaussian approximations to the full metrological sensitivity~\cite{gessner2019metrological,fadel2025}. Their experimental convenience has led to their widespread use, most notably in the form of spin-squeezing coefficients, which play a central role in precision measurements with collective spin systems and provide directly accessible signatures of quantum-enhanced performance and multipartite entanglement~\cite{sorensen2001,sorensen2001extreme,gross2012spin,gross2010,riedel2010,fadel2020,ma2011,pezze2018}.

Squeezing-based approaches are not restricted to Gaussian regimes. By incorporating higher-order moments, generalized squeezing coefficients have been developed that yield substantially tighter bounds for complex, non-Gaussian states~\cite{gessner2019metrological}, while remaining experimentally manageable~\cite{xu2022superconducting}. Importantly, these bounds do not imply a fundamental loss of information: when the optimal observable is accessible, squeezing coefficients recover the full QFI.

Recently it was recognized that the QCRB represents only the lowest element of an entire hierarchy of variance bounds, obtained by imposing increasingly stringent unbiasedness constraints---including constraints at displaced parameter values---each defining a generalized metrological sensitivity~\cite{gessner2023hierarchies,gebhart2024,reichert2026barankin}. Such finite-displacement constructions extend the notion of local unbiasedness around the working point to finite parameter separations and therefore remain applicable even in models with non-differentiable parameter dependence. Other extensions are obtained by imposing higher-order local unbiasedness conditions involving higher-order derivatives. Alternative approaches such as Bayesian bounds and finite-sample analyses also reveal limitations of local QFI-based precision benchmarks~\cite{tsang2012zivzakai,lu2016weissweinstein,rubio2018nonasymptotic,hervas2025corrections,shi2026global}. While the hierarchy introduced in Ref.~\cite{gessner2023hierarchies} extends the notion of quantum limits well beyond the standard framework, practical methods to bound and characterize these generalized sensitivities are so far unexplored. This raises the central open question of whether squeezing-type approaches can be extended from the QFI to the full hierarchy.

Here we introduce squeezing-type lower bounds for the entire hierarchy of generalized metrological sensitivities, which generalize conventional squeezing coefficients. We show how observables optimizing these bounds can be identified analytically within families of accessible observables. If arbitrary observables are accessible, they recover the exact generalized metrological sensitivity, thereby extending the reach of squeezing-based methods beyond the standard QFI.

This article is structured as follows. In Sec. \ref{sec:metrological_sensitivity}, we review the standard framework of quantum parameter estimation and introduce the main ingredients of quantum metrology. We then extend this framework beyond the QCRB by introducing the hierarchy of generalized metrological sensitivities associated with the Barankin, Bhattacharyya, and Abel bounds. In Sec. \ref{sec:quotient_bounds}, we present our main results. We introduce squeezing-type parameters as experimentally accessible lower bounds on these generalized sensitivities and derive their optimization over restricted families of observables. In Sec. \ref{sec:examples_quotient} we illustrate the resulting framework through analytical and numerical examples, including single-qubit probes and collective spin systems undergoing one-axis-twisting dynamics. Finally, we summarize our results and discuss their implications in Sec. \ref{sec:conclusions}.

\section{Metrological sensitivity}\label{sec:metrological_sensitivity}
\subsection{Quantum Cramér-Rao bound and quantum Fisher information}
We are concerned with a quantum system whose state $\rho(\theta)$ depends on the parameter of interest $\theta$ whose precise value is unknown. To infer the value of $\theta$, repeated measurements are performed on $\rho(\theta)$, and an estimator function $\theta_{\text{\rm est}}(x)$ maps the measurement outcomes $x$ to the parameter space. The measurement setting is described by a positive operator-valued measure (POVM), i.e. a set of operators that satisfy $E_x\geq0$ and $\sum_xE_x=\mathds{I}$ for discrete outcomes. For a given observable, the probability to obtain the outcome $x$ given that the parameter of interest is $\theta$ is given by the Born rule, 
\begin{align}
p(x|\theta)=\Tr\{\rho(\theta)E_x\}.    
\end{align}
The precision of the estimator can be quantified by its variance 
\begin{align}\label{eq:estvar}
(\Delta\theta_{\text{\rm est}})^2 = \sum_x p(x|\theta)(\theta_{\text{\rm est}}(x)-\langle\theta_{\text{\rm est}}\rangle_\theta)^2,    
\end{align}
where $\langle\theta_{\text{\rm est}}\rangle_\theta=\sum_xp(x|\theta)\theta_{\text{\rm est}}(x)$ denotes the average value. Let \(\Theta\subset\mathds R\) denote the allowed parameter range. We evaluate the variance at \(\theta\in\Theta\) while additional unbiasedness constraints will introduce a family of bounds. An estimator unbiased locally at \(\theta_0\) has \(\langle\theta_{\rm est}\rangle_{\theta_0}=\theta_0\) for some \(\theta_0\in\Theta\); in particular, it satisfies
\begin{align}\label{eq:local_unb}
\left.\frac{\partial\langle\theta_{\text{\rm est}}\rangle_\theta}{\partial\theta}\right|_{\theta=\theta_0}=1.    
\end{align}
The variance~(\ref{eq:estvar}) is fundamentally limited by the Cramér-Rao bound~\cite{frechet1943,darmois1945,rao1945,cramer1946},
\begin{align}\label{eq:CRB}
    (\Delta\theta_{\text{\rm est}})^2\geq (\Delta\theta_{\text{\rm est}})_{\text{CR}}^2\equiv\frac{1}{F(\theta)},
\end{align}
determined by the Fisher information~\cite{fisher1922},
\begin{align}
    F(\theta)=\sum_x\frac{1}{p(x|\theta)}\left( \frac{\partial p(x|\theta)}{\partial\theta} \right)^2,
\end{align}
which sets a threshold for the achievable precision of the estimator for the chosen measurement.

By optimizing the Fisher information over all POVMs, one obtains the QFI~\cite{Braunstein1994}, 
\begin{align}
    F_Q[\rho(\theta)]=\max_{\{E_x\}_x}F(\theta)=\Tr\{\rho(\theta)L^2_\theta\},
\end{align}
with $L_\theta$ being the symmetric logarithmic derivative (SLD), defined through
\begin{align}
    \frac{\partial \rho(\theta)}{\partial \theta}=\frac{1}{2}(L_{\theta}\rho(\theta)+\rho(\theta)L_{\theta}).
\end{align}
The QFI sets a fundamental, measurement-independent quantum limit on the achievable precision, known as the QCRB, for any estimator satisfying the local unbiasedness condition~\eqref{eq:local_unb}. This bound represents the ultimate lower limit on the variance of any such estimator for the parameter of interest,
\begin{align}\label{eq:QCRB}
    (\Delta\theta_{\text{\rm est}})^2\geq (\Delta\theta_{\text{\rm est}})_{\text{CR}}^2\geq(\Delta\theta_{\text{\rm est}})_{\text{QCR}}^2\equiv\frac{1}{F_Q[\rho(\theta)]}.
\end{align}
Since the Cramér-Rao bound~(\ref{eq:CRB}) can in principle be saturated by the maximum likelihood estimator, the ultimate sensitivity limit expressed by the QCRB~(\ref{eq:QCRB}) represents an achievable precision limit. However, in practice this saturation may only occur after a large number of repeated measurements have been performed. In the finite-data scenario, the QCRB may overestimate the achievable sensitivity of the state and a generalized approach is required to identify stricter sensitivity bounds in this regime.

\subsection{Generalized quantum bounds and information functions}
A generalized framework for obtaining sensitivity bounds beyond the QCRB was introduced in Ref.~\cite{gessner2023hierarchies}. It is based on a choice of a family of test observables $\bm{G}=(G_1,...,G_n)^\top$ that, typically, are functions of the quantum state of the system $\rho(\theta)$. Upon choosing a POVM $\{E_x\}_x$, a classical counterpart to the test observables is found through 
\begin{align}
g_k(x)=\Tr{E_xG_k}.
\end{align}
Associated with these test functions, we can impose bias constraints $\bm{\lambda}=(\lambda_1,\dots,\lambda_n)^\top$ on the estimator, with elements 
\begin{align}\label{eq:unbconst}
\lambda_k=\sum_{x\in X_+}g_k(x)(\theta_{\mathrm{\rm est}}(x)-\langle\theta_{\mathrm{\rm est}}\rangle_\theta),
\end{align}
where $X_+$ are those values of $x$ for which $p(x|\theta)>0$.

Any estimator satisfying the unbiasedness constraints~(\ref{eq:unbconst}) must obey the following sensitivity bound:
\begin{align}
    (\Delta\theta_{\mathrm{\rm est}})^2 \geq (\Delta\theta_{\mathrm{\rm est}})^2_C=\frac{(\bm{a}^\top \bm{\lambda})^2}{\bm{a}^\top \bm{C}\bm{a}},  \label{eq:C-Sch_classical}
\end{align}

where we have introduced the $n\times n$ classical information matrix $\bm{C}$ with elements
\begin{align}
    C_{kl} &= \sum_{x\in X_+}\frac{g_k(x)g_l(x)}{p(x|\theta)},\label{eq:c_classical_matrix}
\end{align}
and $\bm{a}\in\mathds{R}^n$ is an arbitrary vector of coefficients. The inverse of $(\Delta\theta_{\mathrm{\rm est}})^2_C$ defines a classical information function
\begin{align}\label{eq:Cinformation_function}
    I_C[\rho(\theta),T,\{E_x\}_x]=\frac{\bm{a}^\top \bm{C}\bm{a}}{(\bm{a}^\top\bm{\lambda})^2} = \sum_{x\in X_+}\frac{\Tr\{E_x T\}^2}{\Tr\{E_x\rho(\theta)\}}
\end{align}
that acts as a generalized measure of sensitivity of the measurement data to variations under $\theta$, where we have introduced 
\begin{align}\label{eq:Tdef}
    T=\frac{\bm{a}^\top\bm{G}}{\bm{a}^\top \bm{\lambda}},
\end{align}
which itself also depends on the family of states $\rho(\theta)$.

The most stringent restriction on the estimator variance is obtained by maximizing the coefficient vector $\bm{a}$ in Eq.~(\ref{eq:C-Sch_classical}), which leads to
\begin{align}
    \sup_{\bm{a}}\frac{(\bm{a}^\top \bm{\lambda})^2}{\bm{a}^\top\bm{C}\bm{a}} = \bm{\lambda}^\top\bm{C}^{-1}\bm{\lambda},
\end{align}
when $\bm{C}$ is invertible. This bound is achieved by $\bm{a}_{\mathrm{opt}}=\alpha\bm{C}^{-1}\bm{\lambda}$ with $\alpha\in\mathds{R}$ arbitrary.

These classical bounds depend on the chosen measurement setting. Maximum information about the parameter $\theta$ is obtained from the optimal measurement that minimizes the bound on the estimator variance. This defines the corresponding quantum bounds for any choice of $\bm{a}$ as 
\begin{align}\label{eq:genQvariance}
    (\Delta \theta_{\mathrm{\rm est}})^2_Q = \inf_{\{ E_x \} } (\Delta\theta_{\mathrm{\rm est}})^2_C = \frac{(\bm{a}^\top \bm{\lambda})^2}{\sup_{\{ E_x \}}\bm{a}^\top\bm{C}\bm{a}}=\frac{(\bm{a}^\top \bm{\lambda})^2}{\bm{a}^\top\bm{Q}\bm{a}}.
\end{align}
In the last step, we used the fact that this optimization can be determined analytically and leads to the $n\times n$ quantum information matrix $\bm{Q}$ with elements~\cite{gessner2023hierarchies}
\begin{align}
    Q_{kl} = \text{Tr}\{G_k\Omega_{\rho(\theta)}(G_l)\}, \label{eq:q_matrix}
\end{align}
and the super-operator $\Omega_{A}$ known as the symmetric division operator and defined as
\begin{align}\label{eq:Omegadef}
    X=\frac{1}{2}\left(A\Omega_{A}(X)+\Omega_{A}(X)A\right).    
\end{align}
Intuitively, $\Omega_{A}(X)$ describes the inverse of the symmetric multiplication of $X$ by the non-singular operator $A$. 

Again, an optimization over the coefficients $\bm{a}$ leads to the tightest possible bound and determines the quantum limit
\begin{align}\label{eq:genBound}
(\Delta\theta_{\mathrm{\rm est}})^2\geq \sup_{\bm{a}}(\Delta\theta_{\mathrm{\rm est}})^2_Q=\sup_{\bm{a}}\frac{(\bm{a}^\top\bm{\lambda})^2}{\bm{a}^\top\bm{Q}\bm{a}} =\bm{\lambda}^\top \bm Q^{-1}\bm{\lambda},
\end{align}
where saturation is achieved by all $\bm{a}_{\mathrm{opt}}=\alpha\bm{Q}^{-1}\bm{\lambda}$ with $\alpha\in\mathds{R}$. Any estimator that satisfies the constraints $\bm\lambda$ will have Eq.~\eqref{eq:genBound} as the associated quantum limit on its precision.

For any fixed choice of $\bm{a}$, the inverse of the generalized quantum bounds $(\Delta\theta_{\mathrm{\rm est}})_Q^2$, given in Eq.~(\ref{eq:genQvariance}), leads to a generalized quantum information function that is related to its classical counterpart~(\ref{eq:Cinformation_function}) by optimization over all POVMs,
\begin{align}\label{eq:information_function}
     I_{Q} [\rho (\theta),T]&=\sup_{\{E_x\}_x}I_C [\rho (\theta),T,\{E_x\}_x]\notag\\&= \frac{\bm{a}^\top\bm{Q}\bm{a}}{(\bm{a}^\top\bm{\lambda})^2} = \Tr\{\rho(\theta)\Omega^2_{\rho(\theta)}(T)\}.
\end{align}
The optimal measurement that achieves the maximum in Eqs.~(\ref{eq:genQvariance}) and~(\ref{eq:information_function}) can be implemented by a projective measurement in the eigenbasis of $\Omega_{\rho(\theta)}(T)$~\cite{gessner2023hierarchies}.

It is straightforward to verify that the above derivation reproduces the QCRB in Eq.~(\ref{eq:genBound}) and the QFI in Eq.~(\ref{eq:information_function}) when only a single test observable 
\begin{align}\label{eq:G1CRB}
G_1=\frac{\partial \rho(\theta)}{\partial \theta}    
\end{align}
is chosen with unbiasedness condition $\lambda_1=1$. Different choices of test observables yield generalizations of the QCRB including quantizations of the classical bounds first studied by Bhattacharyya~\cite{Bhattacharyya1946}, Barankin~\cite{Barankin1949} or Abel~\cite{Abel1993}, as we will briefly summarize in the next subsection.

If more than one test observable is involved, the bounds depend on the coefficients $\bm{a}$ that determine their linear combination through the definition of the operator $T$ in Eq.~(\ref{eq:Tdef}). In this case, assuming invertibility of the matrix $\bm{Q}$, an optimal choice of $\bm{a}$ and fixed $\bm\lambda$ leads to the coefficient-optimized quantum information function
\begin{align}\label{eq:IQopt}
    I_{Q} [\rho (\theta),\bm{G}]&=\min_{\bm{a}}I_{Q} [\rho (\theta),T]\notag\\&=
    I_{Q} [\rho (\theta),T_{\mathrm{opt}}]\notag\\&=\frac{1}{\bm{\lambda}^\top \bm Q^{-1}\bm{\lambda}} = \Tr\{\rho(\theta)\Omega^2_{\rho(\theta)}(T_{\mathrm{opt}})\},
\end{align}
with
\begin{align}\label{eq:aopt}
    T_{\mathrm{opt}}=\frac{\bm{a}_{\mathrm{opt}}^\top\bm{G}}{\bm{a}_{\mathrm{opt}}^\top \bm{\lambda}},\qquad \bm{a}_{\mathrm{opt}}=\alpha\,\bm{Q}^{-1}\bm{\lambda},
\end{align}
and an irrelevant scaling constant $\alpha\in\mathds{R}$. 

\subsection{Quantum Bhattacharyya, Barankin and Abel bounds}
There are two straightforward ways to extend the approach based on the single test observable~(\ref{eq:G1CRB}) leading to the QFI. The first is to consider test observables $\rho(\theta_{k})$ at finitely shifted parameter values $\theta_{k}=\theta+\lambda_{k}$; this was first proposed for classical probabilities by Barankin~\cite{Barankin1949}. The second is to include higher-order derivatives $G_{k} = \frac{\partial^{k} \rho}{\partial\theta^{k}}$, first pursued in the classical setting by Bhattacharyya~\cite{Bhattacharyya1946}. Both generalizations reproduce the QFI in appropriate limits, either by taking infinitesimal displacements $\lambda_{k}$ or by restricting the construction to the first derivative only. Quantum finite-difference and higher-derivative estimation bounds were previously investigated in Refs.~\cite{TsudaMatsumoto2005,Tsuda2007}.

A linear combination of these two approaches was first suggested for classical parameter estimation by Abel~\cite{Abel1993}. In the quantum setting, this leads to a family of Abel test observables combining the Barankin [$G_{k} = \rho(\theta_{k})$] and Bhattacharyya [$G_{k} = \frac{\partial^{k} \rho}{\partial\theta^{k}}$] constructions. The quantum Abel bound of order $(r,s)$ contains $r$ Barankin test observables and $s$ Bhattacharyya test observables. It is convenient to include additionally the trivial constraint $\lambda_{0}=0$, since this does not increase the number of optimization variables associated with the Barankin terms while yielding a tighter bound. We therefore consider the family of $1+r+s$ Abel test observables
\begin{align}\label{eq:GAbeldef}
    \bm{\tilde{G}} = \left(\rho(\theta), \rho(\theta_{1}), \ldots, \rho(\theta_{r}), \partial_{\theta}\rho(\theta), \ldots, \frac{\partial^{s}\rho(\theta)}{\partial\theta^{s}} \right)^{\top}.
\end{align}

The corresponding fixed unbiasedness constraints are
\begin{align}\label{eq:lambdaAdef}
    \bm{\tilde{\lambda}} = 0\oplus\bm{\lambda} = (0,\lambda_{1},\ldots,\lambda_{r},1,0,\ldots,0)^{\top},
\end{align}
where $\lambda_{k} = \theta_{k} - \theta$ for the $1+r$ Barankin unbiasedness conditions, while for the Bhattacharyya contribution only the first derivative is associated with $\lambda_{k}=1$, and all higher-order derivatives satisfy $\lambda_{k}=0$, as can be seen from Eq.~(\ref{eq:unbconst}). The quantum Abel information matrix then takes the form
\begin{align}
    \bm{\tilde{Q}}_{\mathrm{A}} =
    \begin{pmatrix}
        1 & \bm{e}^{\top} & \bm{0}^{\top} \\
        \bm{e} & \bm{Q}_{\mathrm{Ba}} & \bm{Q}_{\mathrm{H}} \\
        \bm{0} & \bm{Q}_{\mathrm{H}}^{\top} & \bm{Q}_{\mathrm{Bh}}
    \end{pmatrix}, \label{eq:QA_tilde}
\end{align}
with the $r\times r$ quantum Barankin information matrix $\bm{Q}_{\mathrm{Ba}}$
\begin{align}
    (\bm{Q}_{\mathrm{Ba}})_{kl} = \Tr{\rho(\theta+\lambda_{k})\Omega_{\rho(\theta)}(\rho(\theta+\lambda_{l}))},
\end{align}
the $s\times s$ quantum Bhattacharyya information matrix $\bm{Q}_{\mathrm{Bh}}$
\begin{align}
    (\bm{Q}_{\mathrm{Bh}})_{kl} = \Tr{\partial_{\theta}^{k}\rho(\theta)\Omega_{\rho(\theta)}\left(\partial_{\theta}^{l}\rho(\theta)\right)},
\end{align}
and the $r\times s$ hybrid matrix $\bm{Q}_{\mathrm{H}}$, which combines both types of test observables,
\begin{align}
    (\bm{Q}_{\mathrm{H}})_{kl} = \Tr{\rho(\theta+\lambda_{k})\Omega_{\rho(\theta)}\left(\partial_{\theta}^{l}\rho(\theta)\right)}. \label{eq:QH_element}
\end{align}

Selecting the tightest bound among all possible constraints within this family leads to the quantum Abel bound
\begin{align}
    (\Delta\theta_{\rm est})^{2}_{\mathrm{A}} = \sup_{\lambda_{1},\ldots,\lambda_{r}} \bm{\tilde{\lambda}}^{\top}\bm{\tilde{Q}}_{\mathrm{A}}^{-1}\bm{\tilde{\lambda}}. \label{eq:QA_minusff}
\end{align}

If the matrix $\bm{\tilde{Q}}_{\mathrm{A}}$ is invertible, one obtains the simpler form
\begin{align}
    (\Delta\theta_{\rm est})^{2}_{\mathrm{A}} = \sup_{\lambda_{1},\ldots,\lambda_{r}}\bm{\lambda}^{\top}\bigl(\bm{Q}_{\mathrm{A}}-\bm{f}\bm{f}^{\top}\bigr)^{-1}\bm{\lambda},
\end{align}
with $\bm{Q}_{\mathrm{A}}$ denoting the matrix obtained when the trivial constraint $\lambda_{0}=0$ is not included,
\begin{align}
    \bm{Q}_{\mathrm{A}} =
    \begin{pmatrix}
         \bm{Q}_{\mathrm{Ba}} & \bm{Q}_{\mathrm{H}} \\
        \bm{Q}_{\mathrm{H}}^{\top} & \bm{Q}_{\mathrm{Bh}}
    \end{pmatrix}, \label{eq:QAI}
\end{align}
and $\bm{f}=\bm{e}\oplus\bm{0}$, $\bm{e}=(1,1,...,1)^\top\in \mathds{R}^r$, $\bm{0}\in\mathds{R}^s$.

The Abel bound defines the corresponding Abel information function as
\begin{align}
    I_A[\rho(\theta)]
    &:=
    \inf_{\lambda_1,\ldots,\lambda_r}
    I_A[\rho(\theta);\bm\lambda],
    \label{eq:info_Abel_opt}
\end{align}
with
\begin{align}
   I_A[\rho(\theta);\bm\lambda]
    &:=
    I_Q[\rho(\theta),\tilde{\bm G}(\bm\lambda)]
    \nonumber\\
    &=
    \frac{1}{
        \bm{\lambda}^{\top}
        \bigl(
            \bm{Q}_{\mathrm{A}}
            -
            \bm{f}\bm{f}^{\top}
        \bigr)^{-1}
        \bm{\lambda}
    }.
    \label{eq:info_Abel_lambda}
\end{align}

Throughout the following, the notation $I_A[\rho(\theta)]$ is reserved for the Abel information function obtained after optimization over all test points $\lambda_1,\ldots,\lambda_r$, as defined in Eq.~\eqref{eq:info_Abel_opt}, whereas $I_A[\rho(\theta);\bm\lambda]$ denotes the coefficient-optimized quantum information~\eqref{eq:IQopt} for a fixed choice of constraints $\bm\lambda$. Accordingly, a hierarchy between different families of generalized bounds exists between the optimized information functions~\cite{gessner2023hierarchies} and need not hold for arbitrary non-optimal choices of the displacements. 

From the Abel bound, one recovers the Barankin and Bhattacharyya bounds as the special cases $s=0$ and $r=0$, respectively. As in the Abel case, obtaining the Barankin bound from Eq.~\eqref{eq:QA_minusff} yields
\begin{align}
    (\Delta\theta_{\rm est})^{2}_{\mathrm{Ba}} = \sup_{\lambda_{1},\ldots,\lambda_{r}}\bm{\lambda}^{\top}\bigl(\bm{Q}_{\mathrm{Ba}}-\bm{e}\bm{e}^{\top}\bigr)^{-1}\bm{\lambda},
\end{align}
whereas the Bhattacharyya bound is simply
\begin{align}
    (\Delta\theta_{\rm est})^{2}_{\mathrm{Bh}} = (\bm{Q}_{\mathrm{Bh}})^{-1}_{11}.
\end{align}

The same convention applies to the special cases of Barankin and Bhattacharyya. Whenever different $\bm \lambda$ are present, as it is in the case of Barankin ($s=0$), the corresponding information function is obtained only after minimization over these displacements. For the Bhattacharyya family, corresponding to $r=0$, no displaced test points are present and hence no additional optimization is required.

\section{Squeezing-type lower bounds}\label{sec:quotient_bounds}
The optimal observable that achieves the maximum in the definition of the quantum information function~(\ref{eq:information_function}) may not always be experimentally accessible or even theoretically known. Therefore, we will now focus on the derivation of lower bounds on the quantum information function that can be obtained from any accessible observable $X$. In particular, we are interested in bounds that depend only on the first and second moments of $X$. As special cases, these approximations will then reproduce well-known squeezing parameters that are widely employed as easily accessible Gaussian approximations to the QFI in many-body spin systems~\cite{gessner2019metrological}.

\subsection{General result}
We focus on measurements of the fixed observable $X$ with spectral decomposition $X=\sum_x\lambda_x\Pi_x$. The Cauchy-Schwarz inequality $|\sum_x f_x g_x|^2\leq(\sum_x f^2_x)(\sum_x g^2_x)$ with $f_x =\Tr\{\Pi_x T\}/\sqrt{\Tr\{\rho\Pi_x\}}$ and $ g_x = \lambda_x\sqrt{\Tr\{\rho\Pi_x\}}$ leads to
\begin{align}
    \frac{\abs{\Tr\{XT\}}^2}{\Tr\{\rho X^2\}}\leq \sum_x\frac{\Tr\{\Pi_x T\}^2}{\Tr\{\rho\Pi_x\}}= I_C[\rho,T,\{\Pi_x\}_x]. \label{eq:trace}
\end{align}
The right-hand side is the classical information function~(\ref{eq:Cinformation_function}) obtained from a projective measurement in the eigenbasis of $X$. By virtue of Eq.~(\ref{eq:information_function}), any specific measurement choice expressed by a POVM $\{E_x\}_x$ produces a lower bound on the quantum information function,
\begin{align}
    \quad I_C[\rho,T,\{E_x\}_x]\leq I_Q[\rho,T],
\end{align}
and therefore~(\ref{eq:trace}) establishes a lower bound on $I_Q[\rho,T]$. In summary for any observable $X=\sum_x\lambda_x\Pi_x$, we have
\begin{align}\label{eq:qsqzquant}
    \mu^{-2}[\rho,T,X]\leq I_C[\rho,T,\{\Pi_x\}_x]\leq I_Q[\rho,T],
\end{align}
where we have introduced a generalized moment-based sensitivity parameter
\begin{align}\label{eq:chi-2general}
    \mu^{-2}[\rho,T,X]:=\frac{\abs{\Tr\{XT\}}^2}{\Tr\{\rho X^2\}}.
\end{align}

Saturation of the bound~(\ref{eq:trace}) is achieved by the observable
\begin{align}\label{eq:Xopt}
    X=\sum_{x\in X_+}\frac{\Tr\{\Pi_xT\}}{\Tr\{\rho\Pi_x\}}\Pi_x,
\end{align}
up to a scaling factor. The quantum bound~(\ref{eq:qsqzquant}) is then saturated when the projectors $\{\Pi_x\}_x$ constitute the optimal POVM that leads to saturation in Eq.~(\ref{eq:information_function}), which indeed can be achieved by a projective measurement in the eigenbasis of $\Omega_{\rho}(T)$. We can verify this explicitly by considering the spectral decomposition $\Omega_{\rho}(T)=\sum_x\lambda_x\Pi_x$ in Eq.~(\ref{eq:Omegadef}), leading to
\begin{align}
    \Tr\{\Pi_xT\}=\lambda_x\Tr\{\rho \Pi_x\},
\end{align}
and, hence, $\Omega_{\rho}(T)$ indeed takes on the form of the optimal observable $X$ in Eq.~(\ref{eq:Xopt}). For any other suboptimal observable $X$, the moment-based sensitivity~(\ref{eq:chi-2general}) represents a lower bound on the classical and quantum information functions, as expressed by Eq.~(\ref{eq:qsqzquant}).

\subsection{Displacement of the observable}
A case of special interest is obtained by centering the observable $X$, subtracting its expectation value $\langle X\rangle_{\rho}$ with respect to $\rho$. With this choice, the moment-based sensitivity yields the generalized squeezing parameter
\begin{align}
\chi^{-2}[\rho,T,X]&:=\mu^{-2}[\rho,T,X-\langle X\rangle_{\rho}\mathds{1}]\notag\\
&\:=\frac{|\Tr\{XT\}-\Tr\{T\}\Tr\{X\rho\}|^2}{(\Delta X)^2_{\rho}}.\label{eq:gensqzquot}
\end{align}
Such a spectral shift leaves the eigenprojectors $\{\Pi_x\}_x$ of $X$ unchanged and therefore does not affect the associated classical information function $I_C[\rho,T,\{\Pi_x\}_x]$. Consequently, the hierarchy
\begin{align}\label{eq:qsqzquantS}
    \chi^{-2}[\rho,T,X]\leq I_C[\rho,T,\{\Pi_x\}_x]\leq I_Q[\rho,T],
\end{align}
remains valid.

The numerator of Eq.~(\ref{eq:gensqzquot}) contains a contribution that depends on $\Tr\{T\}$. For test observables of Bhattacharyya type,
\begin{align}
    \bm{G}=\bigl(\partial_{\theta}\rho(\theta),\ldots,\partial_{\theta}^{n}\rho(\theta)\bigr)^{\top},
\end{align}
the corresponding trace vanishes identically,
\begin{align}
    \Tr\{T\}
    =
    \frac{\sum_{k=1}^{n}a_{k}\Tr\{\partial_{\theta}^{k}\rho(\theta)\}}{\bm{a}^{\top}\bm{\lambda}}
    =
    0,
\end{align}
since
\begin{align}\label{eq:TBhzero}
    \Tr\!\left\{\partial_{\theta}^{k}\rho(\theta)\right\}
    =
    \partial_{\theta}^{k}\Tr\!\left\{\rho(\theta)\right\}
    =0.
\end{align}
For Barankin-type test observables, however, and therefore also for the hybrid Abel family, this is not automatic.

We next show that $T$ is traceless for all quantum Abel bounds, including the Barankin and Bhattacharyya bounds as special cases, provided that the coefficients $\bm{a}$ in Eq.~(\ref{eq:Tdef}) are chosen optimally according to Eq.~(\ref{eq:genBound}). We thus consider the test observables $\bm{\tilde{G}}$ and unbiasedness conditions $\bm{\tilde{\lambda}}$ that were defined in Eqs.~(\ref{eq:GAbeldef}) and~(\ref{eq:lambdaAdef}), leading to the corresponding extended information matrix $\bm{\tilde{Q}}_{\mathrm{A}}$ in Eq.~\eqref{eq:QA_tilde}.

Recall that the generalized bound is saturated for
\begin{align}
    \bm{\tilde{a}}_{\mathrm{opt}}=\alpha\,\bm{\tilde{Q}}_{\mathrm{A}}^{-1}\bm{\tilde{\lambda}},
\end{align}
with arbitrary $\alpha\in\mathds{R}$~\cite{gessner2023hierarchies}. Moreover, as shown in Eq.~(\ref{eq:TBhzero}), the Bhattacharyya contributions are traceless. Therefore, for the Abel family, $\Tr\{T\}$ is entirely determined by the Barankin terms,
\begin{align}
    \Tr\{T\}
    =
    \frac{\sum_{i=0}^{r}a_{i}}{\bm{\tilde{a}}^{\top}\bm{\tilde{\lambda}}},
\end{align}
where $i=0$ corresponds to the additional observable $\rho(\theta+\lambda_0)$ with $\lambda_{0}=0$. It thus remains to show that $\sum_{i=0}^{r}a_{i}=0$ for the optimal coefficients.

Using the block form of Eq.~\eqref{eq:QA_tilde},
\begin{align}
    \bm{\tilde{Q}}_{\mathrm{A}}
    =
    \begin{pmatrix}
        1 & \bm{f}^{\top} \\
        \bm{f} & \bm{Q}_{\mathrm{A}}
    \end{pmatrix},
\end{align}
the sum of the Barankin coefficients can be written as
\begin{align}
    \sum_{i=0}^{r}a_{i}
    =
    \begin{pmatrix}
        1 & \bm{f}^{\top}
    \end{pmatrix}
    \bm{\tilde{a}}_{\mathrm{opt}}
    =
    \alpha
    \begin{pmatrix}
        1 & \bm{f}^{\top}
    \end{pmatrix}
    \bm{\tilde{Q}}_{\mathrm{A}}^{-1}
    \begin{pmatrix}
        0\\
        \bm{\lambda}
    \end{pmatrix}.
\end{align}
We now use the blockwise inversion formula,
\begin{align}
    \bm{\tilde{Q}}_{\mathrm{A}}^{-1}
    =
    \begin{pmatrix}
        1+\bm{f}^{\top}(\bm{Q}_{\mathrm{A}}-\bm{f}\bm{f}^{\top})^{-1}\bm{f}
        &
        -\bm{f}^{\top}(\bm{Q}_{\mathrm{A}}-\bm{f}\bm{f}^{\top})^{-1}
        \\
        -(\bm{Q}_{\mathrm{A}}-\bm{f}\bm{f}^{\top})^{-1}\bm{f}
        &
        (\bm{Q}_{\mathrm{A}}-\bm{f}\bm{f}^{\top})^{-1}
    \end{pmatrix},
\end{align}
which yields
\begin{align}
    \sum_{i=0}^{r}a_{i}
    &=
    \alpha
    \begin{pmatrix}
        1 & \bm{f}^{\top}
    \end{pmatrix}
    \begin{pmatrix}
        -\bm{f}^{\top}(\bm{Q}_{\mathrm{A}}-\bm{f}\bm{f}^{\top})^{-1}\bm{\lambda}
        \\
        (\bm{Q}_{\mathrm{A}}-\bm{f}\bm{f}^{\top})^{-1}\bm{\lambda}
    \end{pmatrix}
    \\
    &=
    \alpha\left[
    -\bm{f}^{\top}(\bm{Q}_{\mathrm{A}}-\bm{f}\bm{f}^{\top})^{-1}\bm{\lambda}
    +
    \bm{f}^{\top}(\bm{Q}_{\mathrm{A}}-\bm{f}\bm{f}^{\top})^{-1}\bm{\lambda}
    \right]
    =0.\notag
\end{align}
This shows that, when optimized, the coefficients $a_i$ of the Barankin part sum up to zero. Consequently, $\Tr\{T\}=0$, which proves the claim for the full Abel family, including the Barankin and Bhattacharyya bounds as the special cases with $s=0$ and $r=0$, respectively.

\subsection{Optimization over coefficients}
In the following we often work with optimized coefficients. We introduce the normalized test observable
\begin{align}\label{eq:Topt}
    T_{\mathrm{opt}}=\frac{\bm{\tilde{a}}_{\mathrm{opt}}^{\top}\bm{\tilde{G}}}{\bm{\tilde{a}}_{\mathrm{opt}}^{\top}\bm{\tilde{\lambda}}},
\end{align}
with
\begin{align}\label{eq:TrTopt0}
    \Tr\{T_{\mathrm{opt}}\}=0.
\end{align}
Using the definition in Eq.~\eqref{eq:gensqzquot} and the tracelessness of the coefficient-optimized test operator in Eq.~\eqref{eq:TrTopt0}, we introduce the shorthand
\begin{align}
    \chi^{-2}[\rho,\bm{G},X]
    &:=
    \chi^{-2}[\rho,T_{\mathrm{opt}},X]
    \nonumber\\
    &\:=
    \frac{\abs{\Tr\{XT_{\mathrm{opt}}\}}^{2}}
    {(\Delta X)_{\rho}^{2}}.
    \label{eq:chiGdisplacement}
\end{align}
The general hierarchy in Eq.~\eqref{eq:qsqzquantS} then implies
\begin{align}
    \chi^{-2}[\rho,\bm{G},X]
    \leq
    I_Q[\rho,\bm{G}],
    \label{eq:chiIQhierarchyGeneralaopt}
\end{align}
where
$I_Q[\rho,\bm{G}]=I_Q[\rho,T_{\mathrm{opt}}]$
is the coefficient-optimized quantum information function introduced in
Eq.~\eqref{eq:IQopt}.

We recall that the coefficients $\bm{a}$ are chosen so as to provide the most stringent fundamental bound on the variance, i.e., they minimize the quantum information function $I_Q[\rho,T]$. This does not imply that the same coefficients also minimize the squeezing-type lower bound $\chi^{-2}[\rho,T,X]$ for a given observable $X$; in general, they do not.

\subsection{Optimization over measurements}

We now optimize the squeezing coefficient over measurement observables of the form
\begin{align}\label{eq:Xm}
    X_{\bm{m}}=\sum_{i=1}^Km_{i}M_{i},    
\end{align}
where $M_i\in\bm{M}=(M_{1},\ldots,M_{K})^{\top}$ are chosen from a set of accessible observables $\bm{M}$ which need not be orthonormal or complete. The observable is then characterized by the coefficients $\bm{m}\in\mathds{R}^K$. 

The generalized squeezing coefficient~\eqref{eq:gensqzquot} reads
\begin{align}
    \chi^{-2}[\rho,T,X_{\bm m}]
    &=
    \frac{
        \left|
        \sum_i m_i
        \left(
            \Tr\{M_iT\}
            -
            \Tr\{T\}\langle M_i\rangle_\rho
        \right)
        \right|^2
    }{
        \sum_{i,j}m_i\Gamma_{ij}m_j
    }
    \nonumber\\
    &=
    \frac{
        |\bm m^{\top}\bm D_{\rm c}|^2
    }{
        \bm m^{\top}\bm\Gamma\bm m
    },
    \label{eq:chi_measurement_vector}
\end{align}
where
\begin{align}
    \bm D_{\rm c}
    &=
    \bm D
    -
    \Tr\{T\}\langle\bm M\rangle_\rho,
    \nonumber\\
    \bm{\Gamma}
    &=
    \bm{V}
    -
    \langle\bm{M}\rangle_{\rho}\langle\bm{M}\rangle^{\top}_{\rho},
    \label{eq:DG_vector}
\end{align}
and 
$\bm{D}$ and $\bm{V}$ define a vector and a matrix with respective elements
\begin{align}
    D_{i}=\Tr\{M_{i}T\},
    \qquad
    V_{ij}=\frac{1}{2}\langle \{M_{i},M_{j}\}\rangle_{\rho}.
    \label{eq:d_vector}
\end{align}
Optimization over the coefficients $\bm m$ by the Cauchy--Schwarz inequality yields
\begin{align}
    \chi^{-2}[\rho,T,\bm M]
    &:=
    \max_{\bm m}
    \chi^{-2}[\rho,T,X_{\bm m}]
    \nonumber\\
    &=
    \bm D_{\rm c}^{\top}
    \bm\Gamma^{-1}
    \bm D_{\rm c}
    \nonumber\\
    &=
    \left(
        \bm D
        -
        \Tr\{T\}\langle\bm M\rangle_\rho
    \right)^{\top}
    \bm\Gamma^{-1}
    \left(
        \bm D
        -
        \Tr\{T\}\langle\bm M\rangle_\rho
    \right)
    \nonumber\\
    &\leq
    I_Q[\rho,T].
    \label{eq:chiGbound_opt}
\end{align}
The maximizing coefficients satisfy
\begin{align}
    \bm m_{\rm opt}
    \propto
    \bm\Gamma^{-1}\bm D_{\rm c}.
\end{align}
Accordingly, a centered representative of the maximizing observable is
\begin{align}
    X_{\rm opt}
    &=
    \bm D_{\rm c}^{\top}
    \bm\Gamma^{-1}
    \left(
        \bm M
        -
        \langle\bm M\rangle_\rho\mathds{1}
    \right),
    \label{eq:Xopt_accessible}
\end{align}
up to a nonzero overall factor. An arbitrary multiple of the identity may be added without changing the value of $\chi^{-2}$.

An analogous result can be obtained for the generalized moment-based sensitivity~(\ref{eq:chi-2general})
\begin{align}
    \mu^{-2}[\rho,T,X_{\bm{m}}]
    =
    \frac{
    \left|\sum_{i}m_{i}\Tr\{M_{i}T\}\right|^{2}
    }{
    \sum_{i,j}m_{i}\frac{1}{2}\Tr{\{M_{i},M_{j}\}\rho}m_{j}
    }
    =
    \frac{|\bm{m}^{\top}\bm{D}|^{2}}{\bm{m}^{\top}\bm{V}\bm{m}}.
\end{align}
Here, optimizing over the coefficients $\bm{m}$ yields
\begin{align}\label{eq:muopt}
    \mu^{-2}[\rho,T,\bm M]
    &:=
    \max_{\bm m}
    \mu^{-2}[\rho,T,X_{\bm m}]
    \notag\\
    &=
    \bm D^{\top}\bm V^{-1}\bm D
    \leq
    I_Q[\rho,T].
\end{align}
The last step expresses the hierarchy~(\ref{eq:qsqzquant}).

A direct connection between both optimized coefficients~(\ref{eq:chiGbound_opt}) and~(\ref{eq:muopt}) can be established by adding the identity operator to the set of accessible observables,
\begin{align}
    \bm M_{+}
    =
    \left(
        M_1,\ldots,M_K,\mathds{1}
    \right)^{\top}.
\end{align}
Physically, this extra term merely describes an arbitrary constant shift of the observable eigenvalues. The corresponding extended vectors and moment matrix are
\begin{align}
    \bm{D}_+
    =
    \begin{pmatrix}
        \bm{D}\\
        \Tr\{T\}
    \end{pmatrix},
    \qquad
    \bm{V}_+
    =
    \begin{pmatrix}
        \bm{V} & \langle\bm{M}\rangle_{\rho} \\
        \langle\bm{M}\rangle_{\rho}^{\top} & 1
    \end{pmatrix}.
\end{align}
By blockwise inversion,
\begin{align}\label{eq:Ctilde_blockwise}
    \bm{V}_+^{-1}
    =
    \begin{pmatrix}
        \bm{\Gamma}^{-1}
        &
        -\bm{\Gamma}^{-1}\langle\bm{M}\rangle_{\rho}
        \\
        -\langle\bm{M}\rangle_{\rho}^{\top}\bm{\Gamma}^{-1}
        &
        1+\langle\bm{M}\rangle_{\rho}^{\top}
        \bm{\Gamma}^{-1}\langle\bm{M}\rangle_{\rho}
    \end{pmatrix},
\end{align}
with the covariance matrix $\bm{\Gamma}$ defined in Eq.~(\ref{eq:DG_vector}).
The measurement-optimized moment-based sensitivity can therefore be written as
\begin{align}
    \mu^{-2}[\rho,T,\bm{M}_+]&:=\max_{\bm m_+}
    \mu^{-2}[\rho,T,\bm m_+^T\bm{M}_+]\label{eq:chiG_tilde}\\
    &=
\bm{D}_+^{\top}\bm{V}_+^{-1}\bm{D}_+
     \notag\\
    &=\chi^{-2}[\rho,T,\bm M]
    +
    \Tr\{T\}^{2}.\notag
\end{align}
Thus, the optimized uncentered and centered sensitivities differ only by the measurement-independent contribution $\Tr\{T\}^{2}$.

For the optimized test operator in Eq.~\eqref{eq:Topt}, the tracelessness condition in Eq.~\eqref{eq:TrTopt0} implies
$\bm D_{\rm c}=\bm D$. Consequently,
\begin{align}
    \chi^{-2}[\rho,\bm G,\bm M]
    &:=
    \max_{\bm m}
    \chi^{-2}[\rho,\bm G,X_{\bm m}]
    \nonumber\\
    &\:=
    \bm D^{\top}
    \bm\Gamma^{-1}
    \bm D,
    \label{eq:quotient_chi_opt_m}
\end{align}
and, in this case,
\begin{align}
    \mu^{-2}[\rho,\bm G,\bm M_{+}]:=\mu^{-2}[\rho,T_{\rm opt},\bm M_{+}]
    =
    \chi^{-2}[\rho,\bm G,\bm M].
\end{align}
The hierarchy in Eq.~\eqref{eq:qsqzquantS} therefore yields
\begin{align}
    \chi^{-2}[\rho,\bm G,\bm M]
    \leq
    I_Q[\rho,\bm G].
\end{align}

\subsection{Special cases}
\subsubsection{The case $(0,1)$: standard squeezing parameter and QFI}

Let us first consider the standard case described by the Abel bound of order $(0,1)$, i.e., the first-order Bhattacharyya bound. In this case there is a single test observable, $T=\partial_{\theta}\rho(\theta)$. Equation~\eqref{eq:qsqzquantS} recovers on the left-hand side the standard squeezing parameter that was first introduced by Wineland \textit{et al.} in the context of spin observables~\cite{wineland1992spin, wineland1994squeezed}, while the right-hand side reduces to the QFI. Starting, for convenience, from the centered form in Eq.~\eqref{eq:chiGdisplacement}, we find
\begin{align}\label{eq:chi2W}
    \chi^{-2}_{W}[\rho(\theta),X]&:=\chi^{-2}[\rho(\theta),\partial_{\theta}\rho(\theta),X]\notag\\&\;
    =
    \frac{|\Tr\{X\partial_{\theta}\rho(\theta)\}|^{2}}
    {(\Delta X)^{2}_{\rho(\theta)}}
    =
    \frac{\left|\partial_{\theta}\langle X\rangle\right|^{2}}
    {(\Delta X)^{2}_{\rho(\theta)}},
\end{align}
and
\begin{align}
    I_{Q}[\rho(\theta),\partial_{\theta}\rho(\theta)]
    =
    \Tr\!\left\{
    \rho(\theta)
    \Omega^{2}_{\rho(\theta)}
    \!\left(\partial_{\theta}\rho(\theta)\right)
    \right\}
    =
    F_{Q}[\rho(\theta)].
\end{align}
Thus, the general hierarchy~\eqref{eq:qsqzquant} reduces in this case to
\begin{align}\label{eq:Whierarchy}
    \chi^{-2}_{W}[\rho(\theta),X]
    \leq
    F_{Q}[\rho(\theta)],
\end{align}
which is the standard QFI-based squeezing hierarchy. 

The spin-squeezing coefficient was originally introduced for collective spin observables, where a rotation is generated by the Hamiltonian
$H=S_{\bm n_1}$ according to Schrödinger's equation and an orthogonal spin component $X=S_{\bm n_2}$ is measured, such that $(\bm n_1,\bm n_2,\bm n_3)$ are orthonormal directions in $\mathds{R}^3$. Then
\begin{align}
    \left|\partial_\theta\langle S_{\bm n_2}\rangle\right|
    =
    \left|\langle S_{\bm n_3}\rangle\right|,
\end{align}
and Eq.~\eqref{eq:chi2W} becomes
\begin{align}
    \chi^{-2}_{W}[\rho(\theta),S_{\bm n_2}]
    =
    \frac{|\langle S_{\bm n_3}\rangle^2|}
    {(\Delta S_{\bm n_2})^2}.
\end{align}
The conventional Wineland parameter is defined as
\begin{align}
    \xi^2[\rho(\theta),\bm S_{\bm n_2}]
    =
    \frac{N(\Delta S_{\bm n_2})^2}
    {\langle S_{\bm n_3}\rangle^2},
\end{align}
such that
\begin{align}
    \chi_W^{-2}[\rho(\theta),\bm S_{\bm n_2}]=N\xi^{-2}[\rho(\theta),\bm S_{\bm n_2}].
\end{align}
Equation~\eqref{eq:Whierarchy} therefore extends the conventional
Wineland squeezing hierarchy from collective-spin systems to arbitrary
parameterized quantum states and observables.

\subsubsection{The case $(1,0)$: Hammersley--Chapman--Robbins squeezing}

We next consider the Abel bound of order $(1,0)$, corresponding to the Barankin case with one displaced test point~\cite{Hammersley1950,Chapman1951}. The extended set of test observables is $\bm{\tilde{G}}=\bigl(\rho(\theta),\rho(\theta+\lambda)\bigr)^{\top}$, with unbiasedness vector $\bm{\tilde{\lambda}}=(0,\lambda)^{\top}$. The optimized coefficients are given by $\bm{\tilde{a}}_{\mathrm{opt}}=\alpha\,\bm{\tilde{Q}}_{\mathrm{A}}^{-1}\bm{\tilde{\lambda}}$. In the present case, we obtain from Eq.~(\ref{eq:QA_tilde}),
\begin{align}
    \bm{\tilde{Q}}_{\mathrm{A}}
    =
    \begin{pmatrix}
        1 & 1 \\
        1 & Q_{\mathrm{Ba}}
    \end{pmatrix},
    \qquad
    Q_{\mathrm{Ba}}
    =
    \Tr\!\left\{\rho(\theta+\lambda)
    \Omega_{\rho(\theta)}
    \bigl(\rho(\theta+\lambda)\bigr)\right\}.
\end{align}
Its inverse reads
\begin{align}
    \bm{\tilde{Q}}_{\mathrm{A}}^{-1}
    =
    \frac{1}{Q_{\mathrm{Ba}}-1}
    \begin{pmatrix}
        Q_{\mathrm{Ba}} & -1 \\
        -1 & 1
    \end{pmatrix},
\end{align}
and therefore
\begin{align}
    \bm{\tilde{a}}_{\mathrm{opt}}
    =
    \alpha\,\bm{\tilde{Q}}_{\mathrm{A}}^{-1}
    \begin{pmatrix}
        0\\
        \lambda
    \end{pmatrix}
    =
    \frac{\alpha\lambda}{Q_{\mathrm{Ba}}-1}
    \begin{pmatrix}
        -1\\
        1
    \end{pmatrix}.
\end{align}
Thus, up to an irrelevant overall normalization, the optimal coefficients are $\bm{\tilde{a}}_{\mathrm{opt}}=(-1,1)^{\top}$, yielding the test operator
\begin{align}
    T_{\mathrm{HCR}}(\lambda)
    =
    \frac{\rho(\theta+\lambda)-\rho(\theta)}{\lambda}.
\end{align}
For a fixed displacement $\lambda$, Equation~\eqref{eq:chiGdisplacement} then gives
\begin{align}
    \chi^{-2}_{\mathrm{HCR}}[\rho(\theta),X;\lambda]
    &:=
    \chi^{-2}[\rho(\theta),T_{\mathrm{HCR}}(\lambda),X]
    \notag\\
    &=
    \frac{
    \left|
    \langle X\rangle_{\rho(\theta+\lambda)}
    -
    \langle X\rangle_{\rho(\theta)}
    \right|^{2}
    }{
    \lambda^{2}(\Delta X)^{2}_{\rho(\theta)}
    },\label{eq:chi_HCR_lambda}
\end{align}
and the inverse quantum bound is
\begin{align}
    I_{\mathrm{HCR}}[\rho(\theta);\lambda]
    &:=
    I_Q[\rho(\theta),T_{\mathrm{HCR}}(\lambda)]
    \notag\\
    &=
    \Tr\!\left\{
        \rho(\theta)
        \Omega_{\rho(\theta)}^{2}
        \!\left(
            \frac{
                \rho(\theta+\lambda)-\rho(\theta)
            }{
                \lambda
            }
        \right)
    \right\}.
    \label{eq:IHCR_lambda}
\end{align}
Equivalently, \(\lambda^2 I_{\rm HCR}=Q_{\rm Ba}-1\) is the measurement-optimized Bures quantum \(\chi^2\) divergence between the displaced and reference states~\cite{TemmeVerstraete2015}. For every fixed displacement, the general squeezing hierarchy~(\ref{eq:chiIQhierarchyGeneralaopt}) gives
\begin{align}
    \chi^{-2}_{\rm HCR}
    [\rho(\theta),X;\lambda]
    \leq
    I_{\rm HCR}
    [\rho(\theta);\lambda].
    \label{eq:HCR_fixed_lambda_hierarchy}
\end{align}
The HCR information function is then obtained by optimizing over the displacements $\lambda$,
\begin{align}
    I_{\mathrm{HCR}}[\rho(\theta)]
    &:=
    \inf_{\lambda}
    I_{\mathrm{HCR}}[\rho(\theta);\lambda],
    \label{eq:IHCR_opt}
\end{align}
and the resulting hierarchy is therefore
\begin{align}
    \chi^{-2}_{\mathrm{HCR}}[\rho(\theta),X;\lambda^{\rm HCR}_{\rm opt}]
    \leq
    I_{\mathrm{HCR}}[\rho(\theta)],
\end{align}
where $\chi^{-2}_{\mathrm{HCR}}[\rho(\theta),X;\lambda^{\rm HCR}_{\rm opt}]$ is the expression~\eqref{eq:chi_HCR_lambda} evaluated at the optimal displacement that yields~\eqref{eq:IHCR_opt},
\begin{align}
    \lambda^{\rm HCR}_{\rm opt}=\arg\inf_\lambda I_{\rm HCR}[\rho(\theta);\lambda],
\end{align}
whenever the infimum is attained.
This describes the squeezing-type approximation for the quantum Hammersley--Chapman--Robbins (HCR) bound. This finite-distance comparison between the expectation value of $X$ at the test points $\theta+\lambda$ and $\theta$ reduces continuously to the standard QFI-based squeezing hierarchy in the limit $\lambda\to0$, since
$\bigl(\rho(\theta+\lambda)-\rho(\theta)\bigr)/\lambda\to\partial_{\theta}\rho(\theta)$, and, thus,
\begin{align}
    \lim_{\lambda\to0}
    \chi^{-2}_{\mathrm{HCR}}[\rho(\theta),X;\lambda]
    &=
    \frac{\left|\partial_{\theta}\langle X\rangle_{\rho(\theta)}\right|^{2}}
    {(\Delta X)^{2}_{\rho(\theta)}}
    =
    \chi^{-2}_{W}[\rho(\theta),X],
\end{align}
and
\begin{align}
    \lim_{\lambda\to0}
    I_{\mathrm{HCR}}[\rho(\theta);\lambda]
    &=
    I_{Q}[\rho(\theta),\partial_{\theta}\rho(\theta)]
    =
    F_{Q}[\rho(\theta)].
\end{align}
In the following, the HCR quantities at $\lambda=0$ are understood by continuous extension, such that $T_{\rm HCR}|_{\lambda=0}=\partial_\theta\rho(\theta).$

\subsubsection{The case $(1,1)$}
For the case $(1,1)$, the extended set of test observables is
$\bm{\tilde{G}}=(\rho(\theta),\rho(\theta+\lambda),
\partial_{\theta}\rho(\theta))^{\top}$, with
$\bm{\tilde{\lambda}}=(0,\lambda,1)^{\top}$. The corresponding information matrix is
\begin{align}
    \bm{\tilde{Q}}_{\mathrm{A}}
    =
    \begin{pmatrix}
        1 & 1 & 0 \\
        1 & Q_{\mathrm{Ba}} & Q_{\mathrm{H}} \\
        0 & Q_{\mathrm{H}} & Q_{\mathrm{Bh}}
    \end{pmatrix},
\end{align}
where
\begin{align}
Q_{\mathrm{Ba}}
&=
\Tr\!\left\{\rho(\theta+\lambda)
\Omega_{\rho(\theta)}(\rho(\theta+\lambda))\right\},\notag\\
Q_{\mathrm{Bh}}
&=
\Tr\!\left\{\partial_{\theta}\rho(\theta)
\Omega_{\rho(\theta)}(\partial_{\theta}\rho(\theta))\right\},
\end{align}
and
\begin{align}
Q_{\mathrm{H}}
=
\Tr\!\left\{\rho(\theta+\lambda)
\Omega_{\rho(\theta)}(\partial_{\theta}\rho(\theta))\right\}.
\end{align}
The optimized coefficients are
$\bm{\tilde{a}}_{\mathrm{opt}}
=\alpha\,\bm{\tilde{Q}}_{\mathrm{A}}^{-1}\bm{\tilde{\lambda}}$.
Using the block form of Eq.~\eqref{eq:QA_tilde}, we obtain
\begin{align}\label{eq:aopt_Abel11}
    \bm{\tilde{a}}_{\mathrm{opt}}
    =\frac{\alpha}{(Q_{\mathrm{Ba}}-1)Q_{\mathrm{Bh}}
    -
    Q_{\mathrm{H}}^{2}}
    \begin{pmatrix}
        Q_{\mathrm{H}}-\lambda Q_{\mathrm{Bh}} \\
        \lambda Q_{\mathrm{Bh}}-Q_{\mathrm{H}} \\
        Q_{\mathrm{Ba}}-1-\lambda Q_{\mathrm{H}}
    \end{pmatrix}.
\end{align}
The optimized test operator can be written as
\begin{align}
    T_{\mathrm{opt}}^{(1,1)}
    &=
    \frac{
    \lambda Q_{\mathrm{Bh}}-Q_{\mathrm{H}}}{d_{(1,1)}}
    (\rho(\theta+\lambda)-\rho(\theta))
    +
    \frac{Q_{\mathrm{Ba}}-1-\lambda Q_{\mathrm{H}}}{d_{(1,1)}}
    \partial_{\theta}\rho(\theta)
    \notag\\
&=
    \frac{
    \lambda(\lambda Q_{\mathrm{Bh}}-Q_{\mathrm{H}})}{d_{(1,1)}}
    T_{\mathrm{HCR}}(\lambda)
    +
    \frac{Q_{\mathrm{Ba}}-1-\lambda Q_{\mathrm{H}}}{d_{(1,1)}}
    \partial_{\theta}\rho(\theta),
\end{align}
where $d_{(1,1)}=Q_{\mathrm{Ba}}-1+\lambda^{2}Q_{\mathrm{Bh}}-2\lambda Q_{\mathrm{H}}$.

The associated squeezing coefficient~(\ref{eq:chiGdisplacement}) is
\begin{align}
    &\quad\chi^{-2}_{(1,1)}[\rho(\theta),X;\lambda]
    :=
    \chi^{-2}[\rho(\theta),T_{\mathrm{opt}}^{(1,1)},X]
    \\
    &=
    \frac{
    \left|
    \lambda(\lambda Q_{\mathrm{Bh}}-Q_{\mathrm{H}})
    \Delta_{\lambda}\langle X\rangle
    +
    (Q_{\mathrm{Ba}}-1-\lambda Q_{\mathrm{H}})
    \partial_{\theta}\langle X\rangle_{\rho(\theta)}
    \right|^{2}
    }{
    (d_{(1,1)})^{2}
    (\Delta X)^{2}_{\rho(\theta)}
    },\notag
\end{align}
where
$\Delta_{\lambda}\langle X\rangle
=
[\langle X\rangle_{\rho(\theta+\lambda)}
-\langle X\rangle_{\rho(\theta)}]/\lambda$.
The corresponding quantum information function~(\ref{eq:IQopt}) reads
\begin{align}
    I_{(1,1)}[\rho(\theta);\lambda]
    &:=
    I_{Q}[\rho(\theta),T_{\mathrm{opt}}^{(1,1)}]
    \notag\\
    &=
    \frac{
    (Q_{\mathrm{Ba}}-1)Q_{\mathrm{Bh}}
    -
    Q_{\mathrm{H}}^{2}
    }{
    d_{(1,1)}
    }.
\end{align}

Hence, the hybrid Abel squeezing hierarchy reads
\begin{align}
    \chi^{-2}_{(1,1)}[\rho(\theta),X;\lambda]
    \leq
    I_{(1,1)}[\rho(\theta);\lambda]
\end{align}
for any fixed choice of $\lambda$. After optimization of the displacement, the hierarchy becomes
\begin{align}
    \chi^{-2}_{(1,1)}[\rho(\theta),X;\lambda^{(1,1)}_{\rm opt}]\leq I_{(1,1)}[\rho(\theta)],
\end{align}
with $\lambda^{(1,1)}_{\rm opt}=\arg\inf_\lambda I_{(1,1)}[\rho(\theta);\lambda]$ the optimal displacement that achieves
\begin{align}
    I_{(1,1)}[\rho(\theta)]:=\inf_\lambda I_{(1,1)}[\rho(\theta);\lambda].
\end{align}
This is the simplest hybrid case, combining finite-difference and derivative information in the optimized test operator.

\subsubsection{The case $(0,2)$} For the case $(0,2)$, one considers the second-order Bhattacharyya construction with
$\bm{G}=(\partial_{\theta}\rho(\theta),
\partial_{\theta}^{2}\rho(\theta))^{\top}$ and
$\bm{\lambda}=(1,0)^{\top}$. Note that in the case of purely Bhattacharyya bounds, there is no benefit in including the additional observable $\rho(\theta)$ to the set $\bm{G}$, since it decouples from the derivative test observables and its optimal coefficient vanishes. The information matrix~(\ref{eq:QAI}) is therefore
\begin{align}
    \bm{Q}_{\mathrm{Bh}}
    =
    \begin{pmatrix}
        Q_{11} & Q_{12} \\
        Q_{12} & Q_{22}
    \end{pmatrix},
\end{align}
where
\begin{align}
Q_{11}&=F_{Q}[\rho(\theta)],\quad    Q_{12}
=
\Tr\!\left\{\partial_{\theta}\rho(\theta)
\Omega_{\rho(\theta)}(\partial_{\theta}^{2}\rho(\theta))\right\},\notag\\
Q_{22}
&=
\Tr\!\left\{\partial_{\theta}^{2}\rho(\theta)
\Omega_{\rho(\theta)}(\partial_{\theta}^{2}\rho(\theta))\right\}.
\end{align}

Since $\bm{a}_{\mathrm{opt}}
=\alpha\,\bm{Q}_{\mathrm{Bh}}^{-1}\bm{\lambda}$, one obtains
\begin{align}
    \bm{a}_{\mathrm{opt}}
    =
    \frac{\alpha}{Q_{11}Q_{22}-Q_{12}^{2}}
    \begin{pmatrix}
        -Q_{22} \\
        Q_{12}
    \end{pmatrix}.
\end{align}
Thus, up to normalization,
$\bm{a}_{\mathrm{opt}}\propto(-Q_{22},Q_{12})^{\top}$, and
\begin{align}\label{eq:T_op_02}
    T_{\mathrm{opt}}^{(0,2)}
    =
    \partial_{\theta}\rho(\theta)
    -
    \frac{Q_{12}}{Q_{22}}\,
    \partial_{\theta}^{2}\rho(\theta).
\end{align}
The associated squeezing coefficient~(\ref{eq:chiGdisplacement}) is
\begin{align}
    \chi^{-2}_{(0,2)}[\rho(\theta),X]
    &:=
    \chi^{-2}[\rho(\theta),T_{\mathrm{opt}}^{(0,2)},X]
    \notag\\
    &=
    \frac{
    \left|
    \partial_{\theta}\langle X\rangle_{\rho(\theta)}
    -
    \frac{Q_{12}}{Q_{22}}\,
    \partial_{\theta}^{2}\langle X\rangle_{\rho(\theta)}
    \right|^{2}
    }{
    (\Delta X)^{2}_{\rho(\theta)}
    }.\label{eq:chi_Bh_2}
\end{align}
The corresponding quantum information~(\ref{eq:IQopt}) is
\begin{align}
    I_{(0,2)}[\rho(\theta)]
    &:=
    I_{Q}[\rho(\theta),T_{\mathrm{opt}}^{(0,2)}]
    \notag\\
    &=
    Q_{11}-\frac{Q_{12}^{2}}{Q_{22}}
    =
    \frac{Q_{11}Q_{22}-Q_{12}^{2}}{Q_{22}}.
\end{align}
The second-order Bhattacharyya squeezing hierarchy therefore reads
\begin{align}
    \chi^{-2}_{(0,2)}[\rho(\theta),X]
    \leq
    I_{(0,2)}[\rho(\theta)].
\end{align}
This is the natural second-order extension of the standard squeezing parameter. This approach does not include constraints $\bm\lambda$ that require further optimization. 

These examples illustrate the two principal directions in which the standard squeezing hierarchy can be extended: by incorporating finite displacements, as in the Barankin and Abel constructions, or by including higher derivatives, as in the Bhattacharyya hierarchy.

\subsection{Properties}

\subsubsection{Additivity}\label{sec:quotient_bounds_add}

For $m$ identical and independent copies of the system, the generalized squeezing parameter is additive,
\begin{align}
    \chi^{-2}
    \left[
    \rho(\theta)^{\otimes m},
    T^{(m)},
    X^{(m)}
    \right]
    =
    m\,
    \chi^{-2}
    \left[
    \rho(\theta),
    T,
    X
    \right],
    \label{eq:chi_additivity}
\end{align}
where
\begin{align}
    X^{(m)}
    =
    \sum_{i=1}^{m}X^{[i]},
    \qquad
    X^{[i]}
    =
    \mathds{1}^{\otimes(i-1)}
    \otimes X
    \otimes
    \mathds{1}^{\otimes(m-i)}.
\end{align}
The operator $T^{(m)}$ is obtained from $T$ using the same coefficients and normalization, with displaced states and derivatives replaced by their $m$-copy counterparts.

The product-state structure directly implies
\begin{align}
    \left\langle X^{(m)}\right\rangle_{\rho(\theta)^{\otimes m}}
    =
    m\langle X\rangle_{\rho(\theta)}
\end{align}
and
\begin{align}
    \left(\Delta X^{(m)}\right)^2_{\rho(\theta)^{\otimes m}}
    =
    m(\Delta X)^2_{\rho(\theta)}.
    \label{eq:additive_variance}
\end{align}

For a Barankin-type test observable, one has
\begin{align}
    \Tr\!\left\{
    X^{(m)}
    \rho(\theta+\lambda)^{\otimes m}
    \right\}
    =
    m\Tr\!\left\{
    X\rho(\theta+\lambda)
    \right\}.
\end{align}
For a Bhattacharyya-type test observable,
\begin{align}
    \Tr\!\left\{
    X^{(m)}
    \partial_{\theta}^{k}
    \left[
    \rho(\theta)^{\otimes m}
    \right]
    \right\}
    &=
    \partial_{\theta}^{k}
    \left\langle X^{(m)}\right\rangle_{\rho(\theta)^{\otimes m}}
    \notag\\
    &=
    m\Tr\!\left\{
    X\partial_{\theta}^{k}\rho(\theta)
    \right\}.
\end{align}
Hence, by linearity, the same scaling holds for any Abel test operator,
\begin{align}
    \Tr\!\left\{
    X^{(m)}T^{(m)}
    \right\}
    =
    m\Tr\{XT\}.
    \label{eq:additive_test_overlap}
\end{align}

The trace of the test operator is also independent of the number of copies. Indeed, Barankin contributions satisfy
\begin{align}
    \Tr\!\left\{
    \rho(\theta+\lambda)^{\otimes m}
    \right\}
    =
    1,
\end{align}
whereas Bhattacharyya contributions satisfy
\begin{align}
    \Tr\!\left\{
    \partial_{\theta}^{k}
    \left[
    \rho(\theta)^{\otimes m}
    \right]
    \right\}
    =
    \partial_{\theta}^{k}
    \Tr\!\left\{
    \rho(\theta)^{\otimes m}
    \right\}
    =
    0.
\end{align}
Therefore, $\Tr\!\left\{T^{(m)}\right\}
    =
    \Tr\{T\}$, and together with the scaling of the expectation value and
Eq.~\eqref{eq:additive_test_overlap}, this gives
\begin{align}
    &\Tr\!\left\{
        X^{(m)}T^{(m)}
    \right\}
    -
    \Tr\!\left\{T^{(m)}\right\}
    \left\langle
        X^{(m)}
    \right\rangle_{\rho(\theta)^{\otimes m}}
    \nonumber\\
    &\qquad=
    m\left[
        \Tr\{XT\}
        -
        \Tr\{T\}
        \langle X\rangle_{\rho(\theta)}
    \right].
    \label{eq:additive_centered_overlap}
\end{align}
Together with Eq.~\eqref{eq:additive_variance}, this proves
Eq.~\eqref{eq:chi_additivity}.

For fixed test-operator coefficients and displacements, the centered sensitivity is additive under independent repetitions and additive measurements. This property also holds after optimizing the common measurement direction. However, the information-optimal test operator $T_{\rm{opt}}$, Eq.~(\ref{eq:aopt}), is determined from the generalized quantum information matrix and can depend on the number of copies. Consequently, the generalized squeezing coefficient evaluated at $T^{(m)}_{\rm opt}$ need not equal $m$ times its single-copy value.

\subsubsection{Convexity}

We next study the convexity of the generalized squeezing parameter defined in Eq.~\eqref{eq:gensqzquot}. We consider a convex mixture of parameter-dependent states,
\begin{align}
    \rho(\theta)
    =
    \sum_l p_l\rho_l(\theta),
    \qquad
    p_l\geq 0,
    \qquad
    \sum_l p_l=1,
    \label{eq:convex_state_mixture}
\end{align}
where the probabilities $p_l$ are independent of $\theta$. For a fixed choice of the coefficients and displacements entering the test-operator construction, let $T_l$ denote the test operator associated with $\rho_l(\theta)$. Replacing every displaced state and derivative by the corresponding convex mixture gives
\begin{align}
    \rho(\theta+\lambda_j)
    &=
    \sum_l p_l\rho_l(\theta+\lambda_j),
    \nonumber\\
    \partial_\theta^j\rho(\theta)
    &=
    \sum_l p_l\partial_\theta^j\rho_l(\theta).
\end{align}
Since the construction in Eq.~\eqref{eq:Tdef} is linear in the test observables, the test operator associated with $\rho(\theta)$ satisfies
\begin{align}
    T
    =
    \sum_l p_lT_l.
    \label{eq:convex_test_mixture}
\end{align}

For fixed coefficients, displacements, and normalization, the traces of the test operators are also identical for all components of the mixture. This follows because every displaced density operator has unit trace, whereas every derivative of a density operator is traceless. Consequently,
\begin{align}
    \Tr\{T_l\}
    =
    \Tr\{T\}
\end{align}
for every $l$.

For an arbitrary fixed observable $X$, independent of the states in the mixture, Eqs.~\eqref{eq:convex_state_mixture} and \eqref{eq:convex_test_mixture} therefore imply
\begin{align}
    &\Tr\{XT\}
    -
    \Tr\{T\}\langle X\rangle_{\rho(\theta)}
    \nonumber\\
    &\qquad=
    \sum_l p_l
    \left[
        \Tr\{XT_l\}
        -
        \Tr\{T_l\}
        \langle X\rangle_{\rho_l(\theta)}
    \right].
    \label{eq:convex_centered_overlap}
\end{align}
The variance of $X$ in the mixed state satisfies
\begin{align}
    (\Delta X)^2_{\rho(\theta)}
    &=
    \sum_l p_l
    (\Delta X)^2_{\rho_l(\theta)}
    +
    \sum_l p_l
    \left[
        \langle X\rangle_{\rho_l(\theta)}
        -
        \langle X\rangle_{\rho(\theta)}
    \right]^2
    \nonumber\\
    &\geq
    \sum_l p_l
    (\Delta X)^2_{\rho_l(\theta)}.
    \label{eq:convex_variance}
\end{align}
Applying the Cauchy--Schwarz inequality to the right-hand side of Eq.~\eqref{eq:convex_centered_overlap} gives
\begin{align}
    &\quad\left|
    \sum_l p_l
    \left[
        \Tr\{XT_l\}
        -
        \Tr\{T_l\}
        \langle X\rangle_{\rho_l(\theta)}
    \right]
    \right|^2
    \nonumber\\
    &\leq
    \left[
        \sum_l p_l
        (\Delta X)^2_{\rho_l(\theta)}
    \right]
    \left[
        \sum_l p_l
        \frac{
            \left|
                \Tr\{XT_l\}
                -
                \Tr\{T_l\}
                \langle X\rangle_{\rho_l(\theta)}
            \right|^2
        }{
            (\Delta X)^2_{\rho_l(\theta)}
        }
    \right].
\end{align}
Combining this result with Eq.~\eqref{eq:convex_variance}, we obtain
\begin{align}
    \chi^{-2}[\rho(\theta),T,X]
    \leq
    \sum_l p_l\,
    \chi^{-2}[\rho_l(\theta),T_l,X].
    \label{eq:chi_convexity}
\end{align}
Thus, for any fixed choice of the coefficients and displacements defining the test operator, the centered generalized squeezing parameter is convex under mixtures of parameter-dependent state families. Since Eq.~\eqref{eq:chi_convexity} holds for every fixed observable $X$, convexity is preserved under optimization over a common, state-independent family of accessible measurement observables.

The corresponding result for the moment-based sensitivity~(\ref{eq:chi-2general}) follows directly from the linearity of
$\Tr\{XT\}$ and $\Tr\{\rho(\theta)X^2\}$ under the simultaneous mixtures in Eqs.~\eqref{eq:convex_state_mixture} and \eqref{eq:convex_test_mixture}. The same Cauchy--Schwarz argument yields
\begin{align}
    \mu^{-2}[\rho(\theta),T,X]
    \leq
    \sum_l p_l\,
    \mu^{-2}[\rho_l(\theta),T_l,X].
    \label{eq:mu_convexity}
\end{align}
Unlike the centered result, this inequality does not require the traces $\Tr\{T_l\}$ to coincide.

Although the test operator depends on the state family $\rho(\theta)$, this dependence is linear as long as the coefficients defining it are kept fixed. After optimization over these coefficients, however, $\bm a_{\rm opt}$ generally depends on the state through the corresponding quantum information matrix. Consequently, the optimized test operator associated with $\rho(\theta)$ need not satisfy Eq.~\eqref{eq:convex_test_mixture} in terms of the separately optimized test operators associated with the states $\rho_l(\theta)$. Hence, the above argument does not, in general, establish convexity of the fully coefficient-optimized squeezing parameter.

\section{Examples and applications}\label{sec:examples_quotient}

\subsection{Single qubit}

\subsubsection{General properties}
We first illustrate the squeezing hierarchy for a single qubit. We write the state as
$\rho(\theta)=[\mathds{1}+\bm r_\theta^{\top}\bm\sigma]/2$, where
$\bm r_\theta\in\mathds{R}^3$ is its Bloch vector and
$\bm\sigma=(\sigma_x,\sigma_y,\sigma_z)^{\top}$. A general Hermitian single-qubit test operator can be parametrized as
\begin{align}
    T
    =
    \frac{1}{2}
    \left(
        g_0\mathds{1}
        +
        \bm g^{\top}\bm\sigma
    \right),
    \label{eq:single_qubit_general_T}
\end{align}
where $g_0=\Tr\{T\}$ and the vector $\bm g$ depends on the chosen metrological bound. We consider the complete set of spin components as accessible observables,
$\bm M=\bm\sigma/2$. Using
$\{\sigma_i,\sigma_j\}=2\delta_{ij}\mathds{1}$ and
$\langle\sigma_i\rangle_{\rho(\theta)}=(\bm r_\theta)_i$, the covariance matrix is
\begin{align}
    \bm\Gamma
    =
    \frac{1}{4}
    \left(
        \mathds I
        -
        \bm r_\theta\bm r_\theta^{\top}
    \right).
    \label{eq:single_qubit_covariance}
\end{align}
The vector $\bm{D}_c$ introduced in Eq.~\eqref{eq:DG_vector} has components
\begin{align}
    \Tr\{M_iT\}
    -
    \langle M_i\rangle_{\rho(\theta)}
    \Tr\{T\}
    =
    \frac{1}{2}
    \left[
        (\bm g)_i
        -
        g_0(\bm r_\theta)_i
    \right].
\end{align}
Equation~\eqref{eq:chiGbound_opt} therefore yields
\begin{align}
    \chi^{-2}[\rho(\theta),T,\bm M]
    =
    \left(
        \bm g-g_0\bm r_\theta
    \right)^{\top}
    \left(
        \mathds I
        -
        \bm r_\theta\bm r_\theta^{\top}
    \right)^{-1}
    \left(
        \bm g-g_0\bm r_\theta
    \right).
    \label{eq:single_qubit_chi_general}
\end{align}

For comparison, consider the identity-augmented set
\begin{align}
    \bm M_+
    =
    \left(
        \bm\sigma/2,
        \mathds{1}
    \right)^{\top}.
\end{align}
The relation~\eqref{eq:chiG_tilde} gives
\begin{align}
    \mu^{-2}[\rho(\theta),T,\bm M_+]
    =
    \chi^{-2}[\rho(\theta),T,\bm M]
    +
    g_0^2.
    \label{eq:single_qubit_mu_general}
\end{align}

To compare these expressions with the corresponding quantum information function, we explicitly determine the symmetric division operator defined in Eq.~\eqref{eq:Omegadef}. When applied to $T$, its definition yields
\begin{align}
    T&=\frac{1}{2}\left(\rho(\theta)\Omega_{\rho(\theta)}(T)+\Omega_{\rho(\theta)}(T)\rho(\theta)  \right).
\end{align}
We parametrize $\Omega_{\rho(\theta)}(T)=\alpha\mathds{I}+\bm{b}^\top\bm{\sigma}$ and $T=\frac{1}{2}\left( 
    g_0\mathds{I}
    +
    \bm{g}^\top\bm{\sigma}
    \right)$, and use the Bloch vector representation $\rho(\theta)=\frac{1}{2}\left(\mathds{I} +  \bm{r}^\top_{\theta}\bm{\sigma}\right)$ to identify
\begin{align}
   \frac{1}{2}\left( 
    g_0\mathds{I}
    +
    \bm{g}^\top\bm{\sigma}
    \right) &= \frac{1}{2}\left( 
    (\alpha+\bm{r}_\theta^\top\bm{b})\mathds{I}
    +
    (\bm{b}+\alpha\bm{r}_\theta)^\top\bm{\sigma}
    \right),
\end{align}
which implies 
\begin{align}
    \alpha&= \frac{g_0-\bm{r}_\theta^\top\bm{g}}{1-|\bm{r}_\theta|^2},\qquad
    \bm{b} = \bm{g}-\frac{\bm{r}_\theta}{1-|\bm{r}_\theta|^2}\left(g_0-\bm{r}_\theta^\top\bm{g}\right).
\end{align}
The information function, Eq.~\eqref{eq:information_function}, now yields~\cite{gessner2023hierarchies}
\begin{align}
    I_Q[\rho(\theta),T]&= \Tr\{T\Omega_{\rho(\theta)}(T)\}\notag \\
    &= \frac{1}{2}\Tr\left\{ (g_0\alpha + \bm{g}^\top\bm{b})\mathds{I}  \right\}\notag
    \\
    &\qquad+\frac{1}{2}\Tr\left\{ (g_0\bm{b} + \alpha\bm{g} +i(\bm{g}\times\bm{b}))^\top\bm{\sigma}  \right\}\\
    &= g_0\alpha + \bm{g}^\top\bm{b}\notag\\
    &= |\bm{g}|^2+\frac{(g_0-\bm{r}_\theta^\top\bm{g})^2}{1-|\bm{r}_\theta|^2}.\label{eq:IQ1qubit}
\end{align}

For $|\bm r_\theta|<1$, the Sherman--Morrison identity gives
\begin{align}
    \left(
        \mathds I
        -
        \bm r_\theta\bm r_\theta^\top
    \right)^{-1}
    =
    \mathds I
    +
    \frac{
        \bm r_\theta\bm r_\theta^\top
    }{
        1-|\bm r_\theta|^2
    }.
\end{align}
Using this result in Eq.~\eqref{eq:single_qubit_chi_general}, we obtain
\begin{align}
    \chi^{-2}[\rho(\theta),T,\bm M]
    &=
    \left|
        \bm g-g_0\bm r_\theta
    \right|^2
    +
    \frac{
        \left(
            \bm r_\theta^\top\bm g
            -
            g_0|\bm r_\theta|^2
        \right)^2
    }{
        1-|\bm r_\theta|^2
    }
    \nonumber\\
    &=
    |\bm g|^2
    +
    \frac{
        (g_0-\bm r_\theta^\top\bm g)^2
    }{
        1-|\bm r_\theta|^2
    }
    -
    g_0^2.
\end{align}
Comparison with Eq.~\eqref{eq:IQ1qubit} therefore yields
\begin{align}
    \chi^{-2}[\rho(\theta),T,\bm M]
    =
    I_Q[\rho(\theta),T]
    -
    g_0^2.
    \label{eq:single_qubit_centered_relation}
\end{align}
Together with Eq.~\eqref{eq:single_qubit_mu_general}, this implies
\begin{align}
    \mu^{-2}[\rho(\theta),T,\bm M_+]
    =
    I_Q[\rho(\theta),T].
    \label{eq:single_qubit_general_relation}
\end{align}
Thus, for a single qubit, the identity-augmented moment-based sensitivity $\mu^{-2}[\rho(\theta),T,\bm M_+]$ saturates the corresponding quantum information function for an arbitrary test operator $T$. The generalized squeezing parameter $\chi^{-2}[\rho(\theta),T,\bm M]$ differs from it by $g_0^2=\Tr\{T\}^2$.

For the optimized test operator in Eq.~\eqref{eq:Topt}, the tracelessness condition in Eq.~\eqref{eq:TrTopt0} implies
\begin{align}
    g_0=\Tr\{T_{\rm opt}\}=0.
\end{align}
The generalized sensitivities $\chi^{-2}$ and $\mu^{-2}$ therefore coincide and saturate the optimized quantum information function,
\begin{align}
    \chi^{-2}
    [\rho(\theta),T_{\rm opt},\bm M]
    &=
    \mu^{-2}
    [\rho(\theta),T_{\rm opt},\bm M_+]
    \nonumber\\
    &=
    I_Q[\rho(\theta),T_{\rm opt}]
    \nonumber\\
    &=
    \bm g^{\top}
    \left(
        \mathds I
        -
        \bm r_\theta\bm r_\theta^{\top}
    \right)^{-1}
    \bm g.
    \label{eq:single_qubit_chi}
\end{align}

In particular, in the first-order Bhattacharyya or standard Wineland case,
\begin{align}
    \bm g_W(\theta)
    =
    \partial_\theta\bm r_\theta.
\end{align}
For the Hammersley--Chapman--Robbins construction,
\begin{align}
    \bm g_{\rm HCR}(\theta;\lambda)
    =
    \frac{
        \bm r_{\theta+\lambda}-\bm r_\theta
    }{\lambda}.
\end{align}
Finally, for the Abel construction of order $(1,1)$, the two nontrivial contributions can be combined as
\begin{align}
    \bm g_{(1,1)}(\theta;\lambda)
    =
    \frac{
        a_1\bigl(\bm r_{\theta+\lambda}-\bm r_\theta\bigr)
        +
        a_2\partial_\theta\bm r_\theta
    }{
        a_1\lambda+a_2
    },
    \label{eq:single_qubit_g_abel}
\end{align}
where $a_1$ and $a_2$ define the corresponding optimized test operator, with $a_0=-a_1$ already taken into account.

\begin{figure}[tb]
    \centering
    \includegraphics[width=\linewidth]{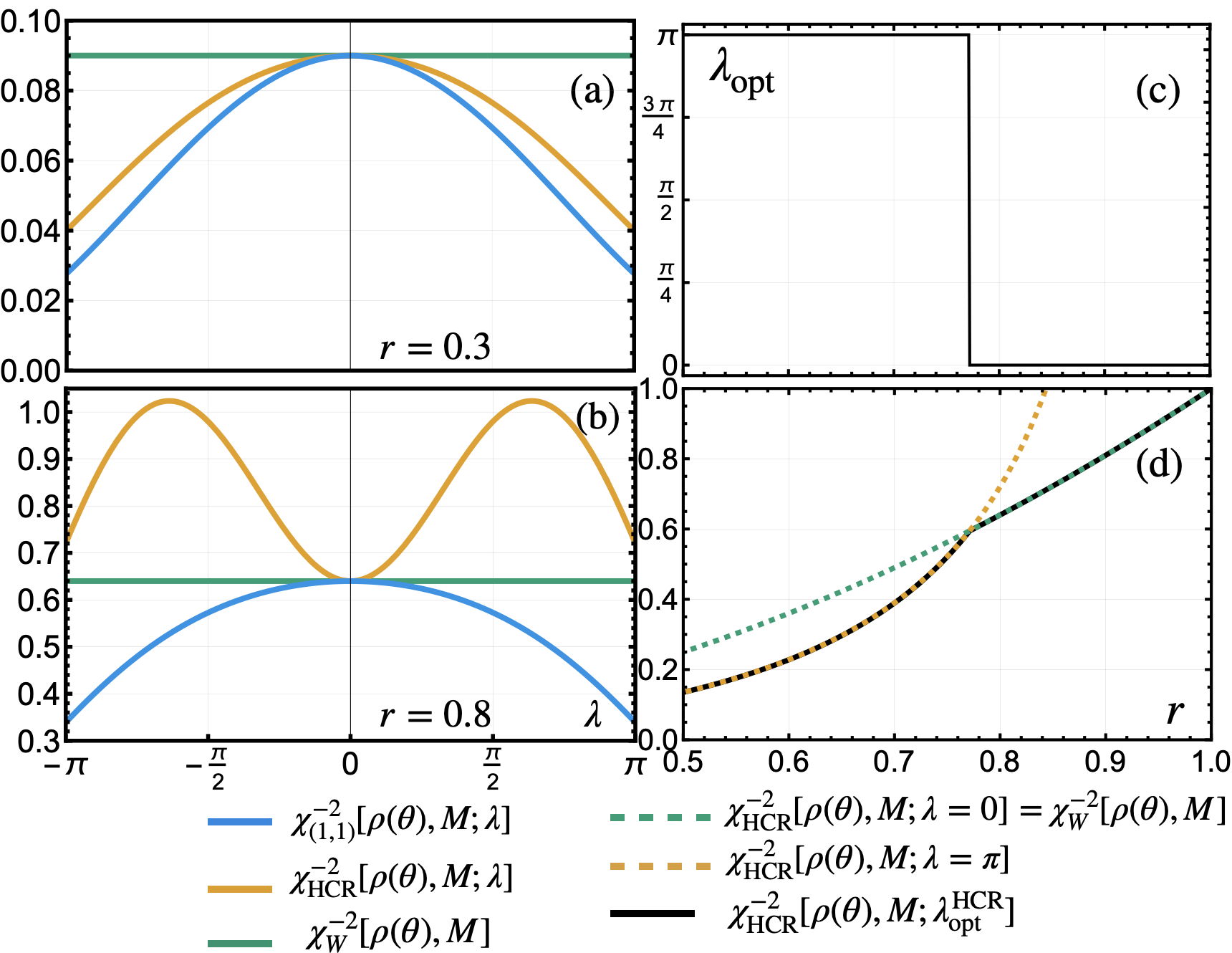}
    \caption{
    Single-copy squeezing-type sensitivities for a single qubit. We compare the Wineland~\eqref{eq:single_qubit_W_explicit}, HCR~\eqref{eq:single_qubit_HCR_explicit}, and Abel~\eqref{eq:single_qubit_Abel_explicit} squeezing sensitivities over $\lambda\in[-\pi,\pi]$ for Bloch-vector lengths (a) $r=0.3$ and (b) $r=0.8$. After optimizing over $\lambda$, the HCR and Abel sensitivities do not exceed the Wineland sensitivity for mixed states. As the state approaches the pure-state limit, the optimized HCR sensitivity converges to the Wineland result, whereas the Abel sensitivity remains lower for all values of $r$ considered here. (c) This behavior can be traced back to the dependence of the optimal displacement $\lambda_{\rm opt}$ on $r$. (d) Above the critical value $r=0.77$ obtained in Eq.~\eqref{eq:single_qubit_critical_r}, the optimal HCR displacement becomes $\lambda_{\rm opt}=0$, thereby recovering the standard Wineland case. 
    }
    \label{fig:single_qubit_r}
\end{figure}

\subsubsection{Example}
We study an explicit comparison of the hierarchy among the different choices of test functions in Fig.~\ref{fig:single_qubit_r}. We consider the initial state
\begin{align}
    \rho_0
    =
    \frac{1}{2}\left(\mathds{1}+r\sigma_y\right),
\end{align}
with $0\leq r<1$. Under the unitary rotation $U_\theta=e^{-i\theta\sigma_z/2}$, the Bloch vector evolves as
\begin{align}
    \bm r_\theta
    =(-r\sin\theta,r\cos\theta,0)^\top.
\end{align}
For the phase-encoding considered here, the sensitivities are invariant under changes of the reference phase $\theta$. Therefore, we may set $\theta=0$ without loss of generality, such that
\begin{align}
    \bm r_0
    =
    \begin{pmatrix}
        0\\ r\\ 0
    \end{pmatrix},
    \qquad
    \left.\partial_\theta\bm r_\theta\right|_{\theta=0}
    =
    \begin{pmatrix}
        -r\\ 0\\ 0
    \end{pmatrix}.
\end{align}
Substitution into Eq.~\eqref{eq:single_qubit_chi} gives the
Wineland and HCR sensitivities
\begin{align}
    \chi_W^{-2}[\rho(\theta),\bm M]
    &=
    r^2 ,
    \label{eq:single_qubit_W_explicit}
    \\
    \chi_{\rm HCR}^{-2}[\rho(\theta),\bm M;\lambda]
    &=
    \frac{r^2}{\lambda^2}
    \left[
        \sin^2\lambda
        +
        \frac{(1-\cos\lambda)^2}{1-r^2}
    \right].
    \label{eq:single_qubit_HCR_explicit}
\end{align}

For the Abel $(1,1)$ case, the coefficients are chosen according to Eq.~\eqref{eq:aopt_Abel11} in order to minimize the corresponding quantum information function. For the present model, the relevant matrix elements reduce to
\begin{align}
    Q_{\rm Bh}
    &=r^2,
    \\
    Q_H
    &=r^2\sin\lambda,
    \\
    Q_{\rm Ba}-1
    &=
    \frac{
        r^2(1-\cos\lambda)
        \left[2-r^2(1+\cos\lambda)\right]
    }{1-r^2}.
\end{align}
Hence, up to an irrelevant common factor, the nontrivial optimal
coefficients can be written as
\begin{align}
    a_1
    &\propto
    \lambda-\sin\lambda ,
    \\
    a_2
    &\propto
    \frac{
        (1-\cos\lambda)
        \left[2-r^2(1+\cos\lambda)\right]
    }{1-r^2}
    -\lambda\sin\lambda .
\end{align}
Inserting these coefficients into
Eq.~\eqref{eq:single_qubit_g_abel} yields
\begin{align}
    &\qquad\chi_{(1,1)}^{-2}[\rho(\theta),\bm M;\lambda]\label{eq:single_qubit_Abel_explicit}\\&
    =
    \frac{
        r^2(1-\cos\lambda)^2
    }{
        (1-\cos\lambda)
        \left[2-r^2(1+\cos\lambda)\right]
        +(1-r^2)
        \left(\lambda^2-2\lambda\sin\lambda\right)
    } .
    \notag
\end{align}

As we have observed above in~(\ref{eq:single_qubit_general_relation}) and~\eqref{eq:single_qubit_chi}, for the present case of a single qubit, the squeezing parameter $\mu^{-2}[\rho(\theta),T,\bm M_+]$ saturate the corresponding quantum information functions for any fixed choice of the coefficients $\bm a$ defining the test operator, while $\chi^{-2}[\rho(\theta),T_{\rm opt}, \bm{M}]$ saturates it for the corresponding $\bm a_{\rm opt}$ that yields a traceless test operator $T$. 
Moreover, all three sensitivities recover the standard result in the local
limit,
\begin{align}
    \lim_{\lambda\rightarrow0}
    \chi_{\rm HCR}^{-2}[\rho(\theta),\bm M;\lambda]
    &=
    \lim_{\lambda\rightarrow0}
    \chi_{(1,1)}^{-2}[\rho(\theta),\bm M;\lambda]\nonumber\\
    &=
    \chi_W^{-2}[\rho(\theta),\bm M]
    =
    r^2 ,
\end{align}
which can be seen in panels $\rm (a)$ and $\rm (b)$ in Fig.~\ref{fig:single_qubit_r}. The abrupt change of the optimal HCR displacement in the right panels in Fig.~\ref{fig:single_qubit_r} can be explained analytically. The two competing minima at $\lambda=0$ and $|\lambda|=\pi$ give
\begin{align}
    \chi_{\rm HCR}^{-2}[\rho(\theta),\bm M;\lambda=0]
    &=r^2,
    \\
    \chi_{\rm HCR}^{-2}[\rho(\theta),\bm M;\lambda=\pi]
    &=
    \frac{4r^2}{\pi^2(1-r^2)}.
\end{align}
Their crossing determines the critical Bloch-vector length
\begin{align}\label{eq:single_qubit_critical_r}
    r_c
    =
    \sqrt{1-\frac{4}{\pi^2}}
    \simeq 0.7712 .
\end{align}
Accordingly, the information-optimal HCR displacement changes from
$|\lambda_{\rm opt}|=\pi$ for $r<r_c$ to
$\lambda_{\rm opt}=0$ for $r>r_c$, where the HCR construction
recovers the standard Wineland result.

\subsubsection{Multiple copies}
When considering multiple copies of a single qubit, $m>1$, the saturation found in Eq.~\eqref{eq:single_qubit_general_relation} and Eq.~\eqref{eq:single_qubit_chi} for $\bm a_{\rm opt}$ is generally lost for the HCR and Abel constructions. For each value of $m$, the test operator is chosen using the coefficients $\bm a_{\rm opt}^{(m)}$ that minimize the corresponding $m$-copy quantum information function, see Ref.~\cite{gessner2023hierarchies}. In general, these coefficients depend on $m$, such that the additivity relation~\eqref{eq:additive_test_overlap} for fixed test-operator coefficients does not directly apply to the fully optimized squeezing parameter.

\begin{figure}
    \centering
    \includegraphics[width=0.9\linewidth]{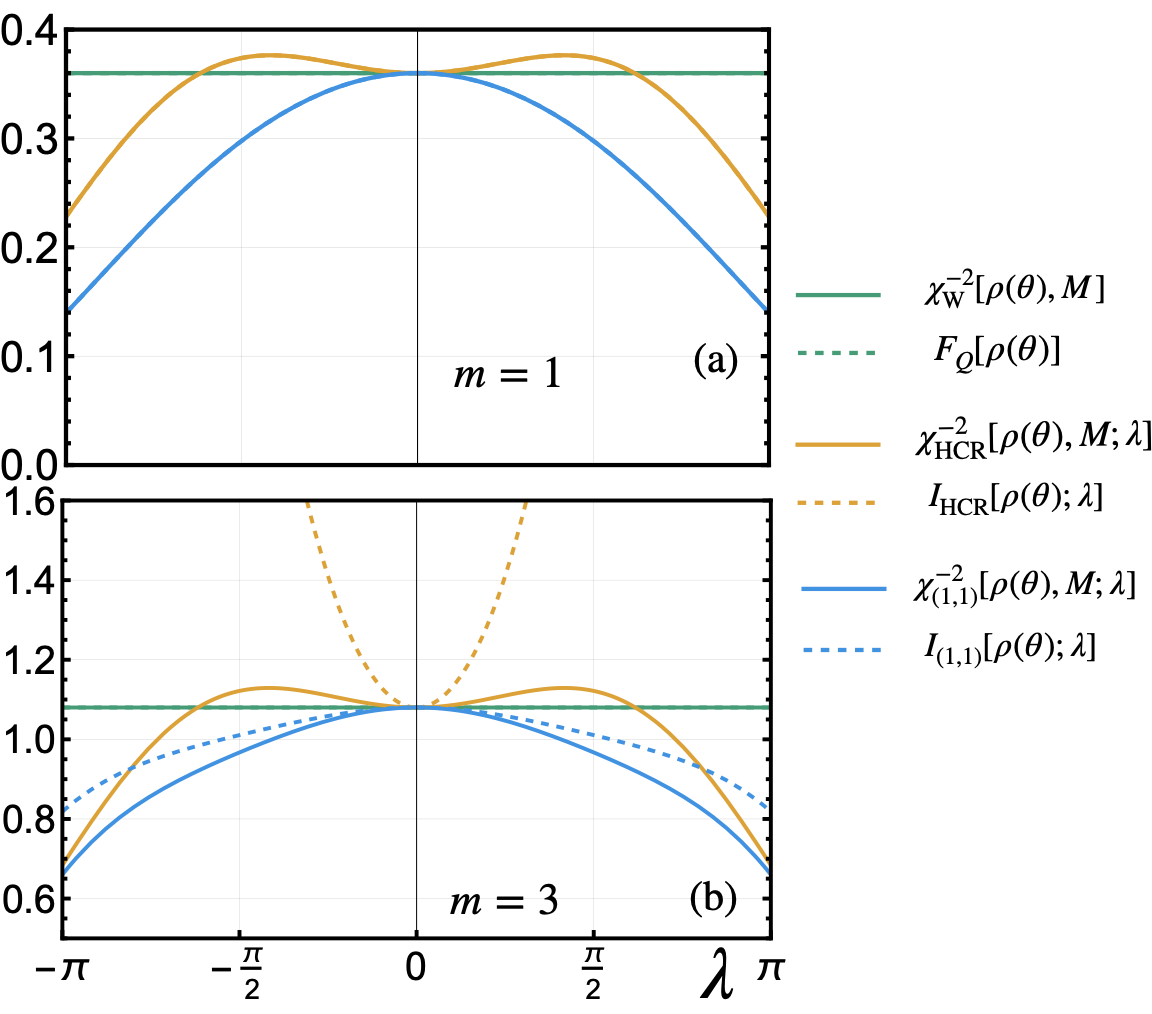}
    \caption{
    Comparison between the squeezing sensitivities and the corresponding quantum information functions for $m=1$ and $m=3$ copies at fixed Bloch-vector length $r=0.6$. Solid lines represent the squeezing-type sensitivities, while dashed lines represent the corresponding quantum information functions for different values of $\lambda$. In each panel, the state and observables appearing in the legend are understood as their corresponding $m$-copy forms. For a single copy, the hierarchy is saturated for all three constructions (a). For multiple copies, saturation is retained only in the standard Wineland--QFI case, while deviations arise for the HCR and Abel constructions. All quantities converge in the limit $\lambda\to0$ (b).
    }
    \label{fig:single_qubit_m}
\end{figure}

For the HCR construction, the optimal coefficients remain proportional to $(-1,1)$ for all $m$, so the corresponding squeezing sensitivity retains its additive form for fixed $\lambda$. For the Abel-$(1,1)$ parameter, the relation $a_0=-a_1$ remains valid, but the relative coefficient $a_2/a_1$ generally depends on $m$, resulting in a non-additive Abel squeezing parameter. Moreover, the associated generalized quantum information functions are themselves not additive in general, and the restricted additive measurement observables considered here no longer necessarily attain them. The standard Wineland-QFI case is the exception, and the additivity of both Wineland squeezing and the QFI ensures saturation of the bound for multiple copies of single-qubits, as is also seen in Fig.~\ref{fig:single_qubit_m}, which compares how these centered sensitivities behave for $m=1$ and $m=3$.

\subsection{Collective phase estimation with $N$ qubits}

We now consider phase estimation with an ensemble of $N$ qubits described by the collective spin operators
$\bm S=(S_x,S_y,S_z)^{\top}$. Restricting to the symmetric subspace with total spin $S=N/2$, the Hilbert-space dimension is $d=N+1$. The system is initially prepared in the $x$-polarized coherent spin state
$|N/2,N/2\rangle_x$, i.e., the eigenstate of $S_x$ with maximal eigenvalue $N/2$.

Spin squeezing is generated by one-axis-twisting (OAT) dynamics~~\cite{kitagawa1993,gross2010,riedel2010,ma2011,pezze2018} with Hamiltonian $H_{\rm OAT}=\hbar\chi S_z^2$, where $\chi$ characterizes the interaction strength. Introducing the dimensionless interaction time
$\tau=\chi t$, the corresponding unitary evolution is
\begin{align}\label{eq:OATevolution}
    U_{\rm OAT}(\tau)
    =
    e^{-i\tau S_z^2},
\end{align}
and the resulting state is
\begin{align}\label{eq:OATstate}
    |\psi_{\rm OAT}(\tau)\rangle
    =
    U_{\rm OAT}(\tau)
    |N/2,N/2\rangle_x.
\end{align}

To model imperfections in the state preparation, we consider the probe state
\begin{align}
    \rho_0
    =
    (1-\epsilon)
    |\psi_{\rm OAT}(\tau)\rangle
    \langle\psi_{\rm OAT}(\tau)|
    +
    \frac{\epsilon}{d}\mathds{1}_d,
    \label{eq:OAT_noisy_state}
\end{align}
where $\epsilon\in[0,1]$ quantifies the admixture of white noise within the symmetric subspace. In the numerical analysis, $\epsilon$ is chosen such that the von Neumann entropy
$S(\rho_0)=-\Tr\{\rho_0\log\rho_0\}$ remains fixed across different values of $N$. The spectrum of $\rho_0$ consists of the nondegenerate eigenvalue $p_0=1-\epsilon+\epsilon/d$ and the $(d-1)$-fold degenerate eigenvalue $p_1=\epsilon/d$. Hence,
\begin{align}
    S(\rho_0)
    =
    -p_0\log p_0
    -
    Np_1\log p_1.
    \label{eq:OAT_entropy}
\end{align}
For a prescribed entropy $S_0$, the corresponding value of $\epsilon$ is therefore $N$-dependent and is obtained numerically from $S(\rho_0)=S_0$. 

The unknown parameter $\theta$ is encoded through the collective unitary transformation
\begin{align}\label{eq:unitary_theta}
    U_\theta=e^{-iH\theta},
    \qquad
    \rho(\theta)=U_\theta\rho_0U_\theta^\dagger,
\end{align}
with the generator
\begin{align}\label{eq:generator}
    H=\bm{S}_n=\frac{1}{2}\sum^N_{i=1}\bm{n}\cdot\bm{\sigma}^{(i)},
\end{align}
and the direction $\bm{n}$ is selected to maximize the QFI of the probe state. The optimal direction is given by the eigenvector associated with the largest eigenvalue of the $3\times3$ covariance matrix, corresponding to the three spin observables $\bm{S}$ of the underlying pure OAT state~(\ref{eq:OATstate}).

We analyze this system using the measurement-optimized squeezing parameter introduced in Eq.~\eqref{eq:quotient_chi_opt_m}, restricting the accessible observables to collective spin components. We compare three representative test-operator constructions. The first is the standard $(0,1)$ case, corresponding to the first-order Bhattacharyya construction and yielding the Wineland spin-squeezing parameter. The second is the first-order Barankin or HCR construction $(1,0)$. Finally, we consider the hybrid Abel construction $(1,1)$, which combines finite-displacement and derivative contributions.

The HCR and Abel constructions require the state to be evaluated both at $\theta$ and at a displaced parameter value $\theta+\lambda$. The displacement range specifies the finite interval over which the additional unbiasedness constraints are imposed. To illustrate a nonlocal but finite estimation problem, we restrict the admissible displacements to the numerical interval $\lambda\in[-2\pi/3,2\pi/3]$ which specifies the allowed displaced test points relative to the reference value \(\theta=0\). Each fixed \(\lambda\) defines its corresponding test-point constraint; the optimization below selects the information-optimal member of this family. In the present phase-estimation setting, both the information functions and the corresponding squeezing parameters are symmetric under $\lambda\to-\lambda$, so the numerical optimization may be restricted to
$\lambda\in[0,2\pi/3]$.

For each OAT time $\tau$, the displacement is chosen according to the corresponding fundamental information function. More precisely, for the HCR case,
\begin{align}
    \lambda_{\rm opt}^{\rm HCR}(\tau)
    =
    \arg\min_{\lambda\in[0,\lambda_{\max}]}
    I_{\rm HCR}[\rho(\theta);\lambda],
    \qquad
    \lambda_{\max}=\frac{2\pi}{3},
    \label{eq:lambdaopt_HCR}
\end{align}
and analogously for the Abel construction. Since the information functions are the inverses of the associated variance bounds, this choice identifies the most restrictive bound within the allowed interval. The squeezing parameters are then evaluated at the corresponding information-optimized displacement.

\begin{figure}[tb]
    \centering
    \includegraphics[width=0.85\linewidth]{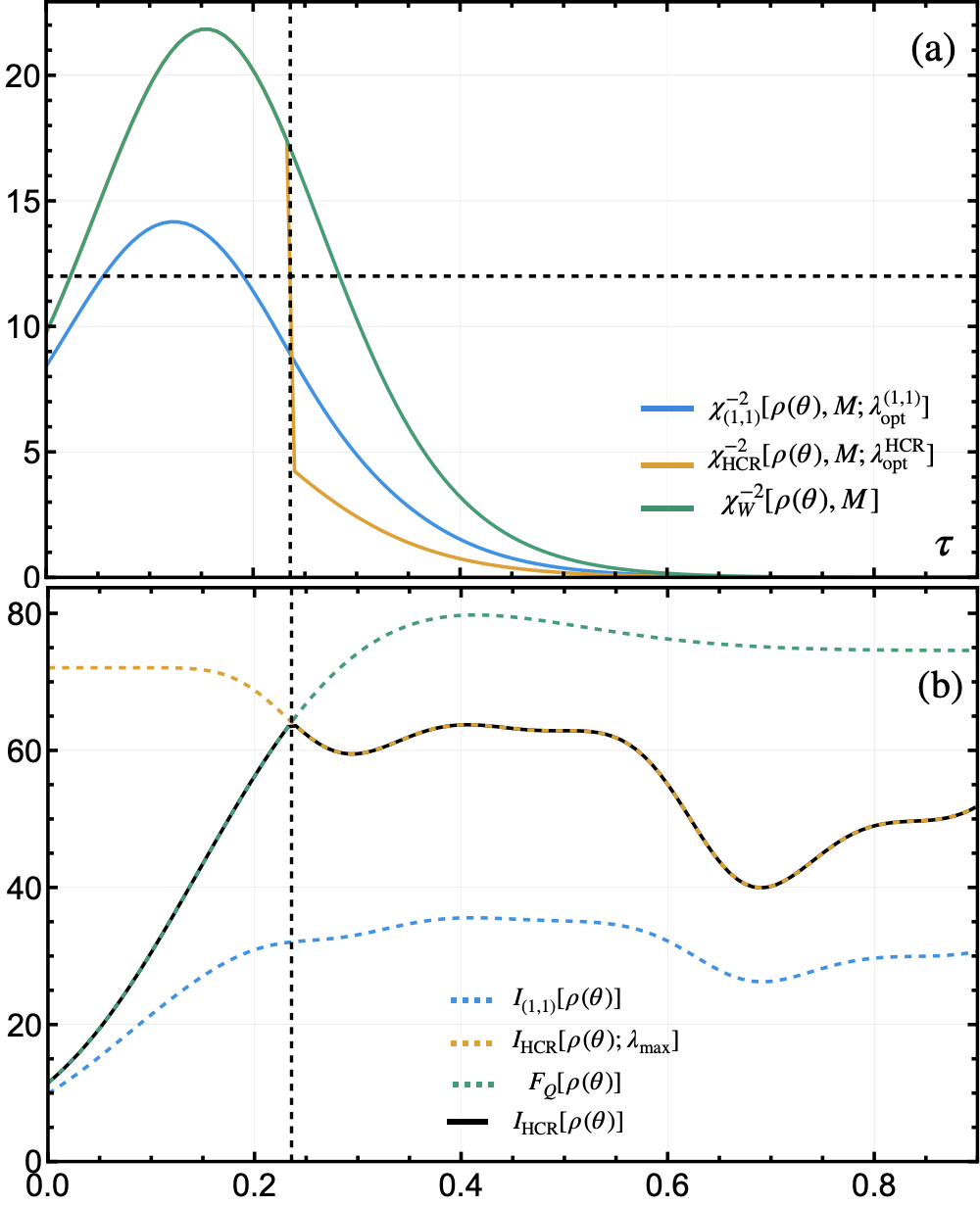}
    \caption{
    Single-copy squeezing parameters for $S(\rho_0)=S_0=0.240$, corresponding to $\epsilon=0.038$ for $N=12$ qubits. The standard Wineland parameter provides the largest sensitivity, while the Abel parameter exceeds the classical limit $N$ (black dashed line) over a finite interval of OAT times (a). At early times, the HCR information function is minimized at $\lambda=0$ and coincides with the QFI. Beyond a critical time, the minimum shifts to $\lambda=\lambda_{\max}=2\pi/3$, producing the abrupt change in the HCR squeezing curve (b). 
    }
    \label{fig:plots_chi}
\end{figure}

In addition to the optimization over $\lambda$, the Abel construction involves the optimization over the coefficients defining the test operator, performed according to Eq.~\eqref{eq:aopt}. We also consider the second-order Bhattacharyya construction $(0,2)$. As shown in Appendix~\ref{sec:appendix}, for the present model the corresponding $(0,2)$ bound reduces exactly to the first-order $(0,1)$ result and therefore provides no additional information. We consequently restrict to the standard, HCR, and Abel constructions.

Figure~\ref{fig:plots_chi} shows the resulting squeezing parameters as functions of the OAT evolution time. For the unitary encoding in Eq.~\eqref{eq:unitary_theta}, we denote the QFI explicitly as $F_Q[\rho_0,H]$ with $H$ the generator Eq.~\eqref{eq:generator}. For any fully separable $N$-qubit state $\rho_{\rm sep}=\sum_kp_k\rho^{(1)}_k\otimes\cdots\otimes\rho^{(N)}_k$, where $\{p_k\}_k$ is a probability distribution and $\rho^{(i)}_k$ are local quantum states of system $i$, collective rotations generated by $H=\bm{S}_n$ satisfy~\cite{pezze2009}
\begin{align}\label{eq:FQ_separability}
    F_Q[\rho_{\rm sep},\bm{S}_n]\leq N,
\end{align}
which defines the usual separability, or shot-noise, threshold for this normalization, such that sensitivities above $N$ certify metrologically useful entanglement. The standard Wineland parameter surpasses this limit over a finite interval and provides the largest sensitivity throughout the evolution, and its corresponding quantum bound, the QFI, serves as an upper reference for the HCR and Abel constructions
\begin{align}
    \chi^{-2}_{\rm HCR}[\rho(\theta),\bm M;\lambda^{\rm HCR}_{\rm opt}] \leq I_{\rm HCR}[\rho(\theta)] &\leq F_Q[\rho(\theta)],\\
    \chi^{-2}_{(1,1)}[\rho(\theta),\bm M;\lambda^{\rm (1,1)}_{\rm opt}] \leq I_{(1,1)}[\rho(\theta)] &\leq F_Q[\rho(\theta)].
\end{align}

The Abel parameter also exceeds $N$ over a finite time interval, showing that the hybrid finite-displacement construction can still certify metrologically useful entanglement and demonstrate sensitivity beyond the shot-noise limit in this setting.

The HCR parameter initially coincides with the Wineland parameter but becomes more conservative at later times. This behavior follows from the optimization over $\lambda$. At early times, the minimum of the HCR information function occurs at $\lambda_{\rm opt}=0$, where the HCR construction reduces to the first-order Bhattacharyya case. Beyond a critical OAT time, the minimum shifts discontinuously to the boundary of the allowed interval,
$\lambda_{\rm opt}=\lambda_{\max}$. This change produces the abrupt transition observed in the HCR squeezing curve. The lower panel of Fig.~\ref{fig:plots_chi} illustrates this mechanism by comparing the HCR information function at $\lambda=\lambda_{\max}$ with the QFI.

\begin{figure}[tb]
    \centering
    \includegraphics[width=0.8\linewidth]{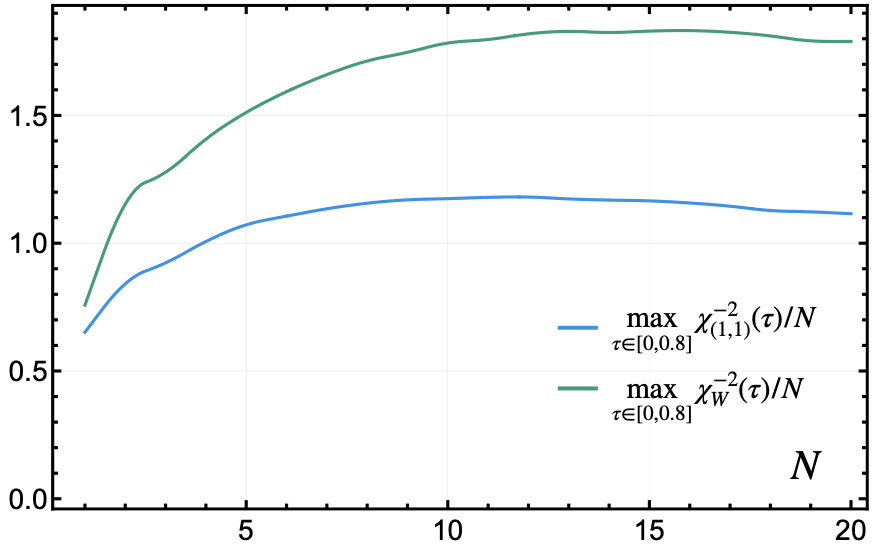}
    \caption{
   Maximum normalized squeezing sensitivity as a function of the particle number $N$ for the Wineland and Abel $(1,1)$ constructions at fixed entropy $S_0=0.24$. The white-noise fraction $\epsilon$ is adjusted for each $N$ according to $S(\rho_0)=S_0$. Although $\epsilon$ decreases with increasing $N$, the metrological gain per particle reaches a maximum and subsequently decreases, reflecting the competition between OAT-generated correlations, normalization by $N$, and noise sensitivity.
    }
    \label{fig:N_scaling}
\end{figure}

To further characterize the system-size dependence, we evaluate the maximum squeezing sensitivity attained by the Abel $(1,1)$ and Wineland constructions for different particle numbers $N$, while keeping the entropy fixed at the same value $S_0=0.24$. The maximization is performed over the OAT interval $\tau\in[0,0.8]$, beyond which both squeezing parameters decrease noticeably for all the system sizes considered here. For small $\epsilon$, Eq.~\eqref{eq:OAT_entropy} gives
\begin{align}
    S(\rho_0)
    =
    \frac{N}{N+1}\epsilon
    \left[
        1-\log\epsilon+\log(N+1)
    \right]
    +O(\epsilon^2).
    \label{eq:entropy_small_epsilon}
\end{align}
Thus, at fixed $S_0$, the white-noise fraction $\epsilon$ decreases with increasing $N$, since the same entropy can be distributed over a larger Hilbert space. This decrease is only logarithmic to leading order and therefore does not guarantee a monotonic improvement of the normalized metrological sensitivity. In Fig.~\ref{fig:N_scaling}, we show $\max_{\tau\in[0,0.8]}\chi^{-2}(\tau)/N$ for the Wineland and Abel $(1,1)$ constructions. The normalization by $N$ measures the metrological gain per particle, whose behavior results from the competition between OAT-generated correlations and their sensitivity to white noise. At constant entropy, this leads to an optimal finite $N$ where the quantum gain is maximized due to the above-described $N$-dependent behavior of the white noise $\epsilon(N)$.

\section{Conclusions}\label{sec:conclusions} 

We introduced squeezing-type parameters that provide accessible lower bounds on the hierarchy of generalized information functions of Ref.~\cite{gessner2023hierarchies}, while retaining the structure of conventional squeezing coefficients. The corresponding sensitivity bound is saturated by an optimal observable. Moreover, optimization of the general extended Abel test operator renders it traceless, yielding a compact covariance-based formulation of the squeezing parameter optimized over accessible measurements.

For fixed test-operator coefficients and displacements, we established additivity for independent copies and convexity under mixtures. The examples of single-qubit probes and collective states generated by one-axis-twisting dynamics illustrate both the attainability of the bounds and the role of finite displacements. For a single qubit, a complete set of spin observables suffices to saturate the corresponding generalized information function, whereas finite-displacement constructions generally lose this saturation property for multiple copies. For the squeezed collective states, the optimized sensitivity can depart from standard local squeezing sensitivity when a nonzero displacement becomes optimal.

These results extend the squeezing approach beyond the quantum Fisher information to a broader family of variance bounds relevant for finite-data and finite-displacement phase estimation. They provide a practical route to accessing generalized metrological sensitivities with restricted measurements, connecting this hierarchy of bounds to experimentally accessible observables.

\appendix

\section{$(0,2)$ squeezing parameter for OAT dynamics}\label{sec:appendix}

We now show that the second-order Bhattacharyya construction $(0,2)$
reduces identically to the first-order $(0,1)$ formulation for the
probe states considered in Sec.~\ref{sec:examples_quotient}. After the unitary parameter encoding in Eq.~\eqref{eq:unitary_theta}, the noisy OAT state Eq.~\eqref{eq:OAT_noisy_state} can be written as
\begin{align}
    \rho(\theta)
    &=
    (1-\epsilon)P_\theta
    +
    \frac{\epsilon}{d}\mathds{1}
    \nonumber\\
    &=
    \left(1-\epsilon+\frac{\epsilon}{d}\right) P_\theta
    +
    \frac{\epsilon}{d} P_\theta^\perp,
    \label{eq:OAT_projector_decomposition}
\end{align}
 for $0<\epsilon\leq1$, where $P_\theta =|\Psi_\theta\rangle\langle\Psi_\theta|$ is the projector onto the pure state $|\Psi_\theta\rangle = U_\theta|\psi_{\rm OAT}\rangle$, $P_\theta^\perp=\mathds{1}-P_\theta$.

For a state consisting of a pure state admixed with white noise, the symmetric division superoperator can be expressed in terms of the two eigenspaces as~\cite{gessner2023hierarchies}
\begin{align}
    \Omega_{\rho(\theta)}(X)
    &=
    \frac{1}{p_0}
    P_\theta X P_\theta
    +
    \frac{2}{p_0+p_1}
    \left(
        P_\theta X P_\theta^\perp
        +
        P_\theta^\perp X P_\theta
    \right)
    \nonumber\\
    &\quad
    +
    \frac{1}{p_1}
    P_\theta^\perp X P_\theta^\perp ,
    \label{eq:Omega_white_noise}
\end{align}
with $p_0 = 1-\epsilon+\frac{\epsilon}{d}$ and $p_1 = \frac{\epsilon}{d}$.

Since $P_\theta$ is a projector, $P_\theta^2=P_\theta$. Differentiating this identity and projecting onto the two eigenspaces gives
\begin{align}
    P_\theta
    (\partial_\theta P_\theta)
    P_\theta
    &=
    0,
    &
    P_\theta^\perp
    (\partial_\theta P_\theta)
    P_\theta^\perp
    &=
    0.
    \label{eq:Pdot_offdiagonal}
\end{align}
Hence, $\partial_\theta P_\theta$ contains only off-diagonal matrix elements between the subspaces associated with $P_\theta$ and $P_\theta^\perp$. Since
\begin{align}
    \partial_\theta\rho(\theta)
    =
    (p_0-p_1)\partial_\theta P_\theta,
\end{align}
the same property holds for $\partial_\theta\rho(\theta)$.

We can now evaluate the off-diagonal element of the second-order Bhattacharyya information matrix,
\begin{align}
    Q_{12}
    =
    \Tr\left\{
        \partial_\theta\rho(\theta)
        \,
        \Omega_{\rho(\theta)}
        \bigl(
            \partial_\theta^2\rho(\theta)
        \bigr)
    \right\}.
    \label{eq:Q12_start}
\end{align}
Because $\partial_\theta\rho(\theta)$ has only off-diagonal blocks with respect to $P_\theta\oplus P_\theta^\perp$, the first and last terms in Eq.~\eqref{eq:Omega_white_noise} do not contribute to the trace. We therefore obtain
\begin{align}
    Q_{12}
    &=
    \frac{2}{p_0+p_1}
    \Tr\left\{
        \partial_\theta\rho(\theta)
        \,
        \partial_\theta^2\rho(\theta)
    \right\}
    \nonumber\\
    &=
    \frac{1}{p_0+p_1}
    \partial_\theta
    \Tr\left\{
        [\partial_\theta\rho(\theta)]^2
    \right\}.
    \label{eq:Q12_HS}
\end{align}

For the unitary family in Eq.~\eqref{eq:unitary_theta}, with a $\theta$-independent generator $H$,
\begin{align}
    \partial_\theta\rho(\theta)
    =
    U_\theta
    \left[
        \partial_\theta\rho(\theta)
    \right]_{\theta=0}
    U_\theta^\dagger.
\end{align}
It follows from unitary invariance of the Hilbert--Schmidt norm that
\begin{align}
    \Tr\left\{
        [\partial_\theta\rho(\theta)]^2
    \right\}
    =
    \Tr\left\{
        [\partial_\theta\rho(0)]^2
    \right\},
\end{align}
and therefore
\begin{align}
    Q_{12}=0.
    \label{eq:Q12_zero_OAT}
\end{align}

Using Eq.~\eqref{eq:Q12_zero_OAT} in Eq.~\eqref{eq:T_op_02}, the optimal test operator reduces to
\begin{align}
    T_{\rm opt}^{(0,2)}
    =
    \partial_\theta\rho(\theta),
\end{align}
while the corresponding quantum information function becomes
\begin{align}
    I_{(0,2)}[\rho(\theta)]
    =
    Q_{11}
    =
    F_Q[\rho(\theta)].
\end{align}
Likewise, the squeezing parameter reduces to the Wineland formulation
\begin{align}
    \chi_{(0,2)}^{-2}[\rho(\theta),X]
    &=
    \frac{
        |\partial_\theta\langle X\rangle_{\rho(\theta)}|^2
    }{
        (\Delta X)^2_{\rho(\theta)}
    }
    \nonumber\\
    &=
    \chi_W^{-2}[\rho(\theta),X].
\end{align}
Thus, for the white-noise-admixed OAT probe and unitary parameter encoding considered here, the second-order Bhattacharyya construction provides no additional sensitivity beyond the standard first-order $(0,1)$ formulation.

\textit{AI usage statement}-- We acknowledge the usage of AI-based tools to assist with language refinement and editing for clarity. The authors also used AI tools to cross-check derivations during preparation of this work. The research results and scientific conclusions were developed by the authors, who take full responsibility for the content.

\textit{Data Availability }-- There are no publicly available research data or software supporting this manuscript. Requests for further information or data should be sent to the authors.

\bibliographystyle{apsrev4-2}
\bibliography{references}

\end{document}